\documentclass{article} % For LaTeX2e
\usepackage{iclr2024_conference,times}

\usepackage{amsmath,amsfonts,bm}

\def\eqref#1{equation~\ref{#1}}
\def\1{\bm{1}}

\DeclareMathAlphabet{\mathsfit}{\encodingdefault}{\sfdefault}{m}{sl}
\SetMathAlphabet{\mathsfit}{bold}{\encodingdefault}{\sfdefault}{bx}{n}

\usepackage{multirow} 
\usepackage{hyperref}
\usepackage{graphicx}
\usepackage{url}

\usepackage{times}
\usepackage{epsfig}
\usepackage{amsmath}
\usepackage{amssymb}
\usepackage{amsthm}
\usepackage{amsfonts,bm}

\usepackage{xpatch}
\usepackage[linesnumbered,ruled]{algorithm2e}
\usepackage[tiny]{titlesec}

\usepackage{booktabs}
\usepackage{bbding}
\usepackage{epigraph}
\usepackage{tcolorbox}
\usepackage{colortbl, xcolor}

\definecolor{grey}{rgb}{0.1,0.1,0.1} 
\usepackage{pifont}

\title{A Comprehensive Review of Community Detection in Graphs}

\iclrfinalcopy

\author{
\textbf{Jiakang Li\thanks{The first two authors contributed equally to this work.}$^{*,1}$, Songning Lai$^{*,2,3}$, Zhihao Shuai$^{5}$, Yuan Tan$^{1}$, Yifan Jia$^{3}$,} \\
\textbf{Mianyang Yu$^{5}$, Zichen Song$^{1}$, Xiaokang Peng$^{1}$, Ziyang Xu$^{4}$, Yongxin Ni$^{6}$,}\\
\textbf{Haifeng Qiu$^{7}$, Jiayu Yang$^{8}$, Yutong Liu$^{6}$, Yonggang Lu\thanks{Correspondence to Yonggang Lu \{ylu@lzu.edu.cn\}.}$^{\dagger,1}$}\\
$^1$School of Information Science and Engineering, Lanzhou University\\
$^2$Hong Kong University of Science and Technology(Guangzhou)\\ 
$^3$School of Information Science and Engineering, Shandong University\\
$^4$School of Mathematics and Statistics, Lanzhou University\\
$^5$School of Big Data and Software Engineering, Chongqing University\\
$^6$National University of Singapore, Singapore\\
$^7$School of Mechanical and Electrical Engineering, Guangzhou University\\
$^8$Hongshen Honors School \& School of Mathematics and Statistics, Chongqing University\\
}

\begin{document}

\maketitle

\begin{abstract}
The study of complex networks has significantly advanced our understanding of community structures which serves as a crucial feature of real-world graphs. Detecting communities in graphs is a challenging problem with applications in sociology, biology, and computer science. Despite the efforts of an interdisciplinary community of scientists, a satisfactory solution to this problem has not yet been achieved. This review article delves into the topic of community detection in graphs, which serves as a thorough exposition of various community detection methods from perspectives of modularity-based method, spectral clustering, probabilistic modelling, and deep learning. Along with the methods, a new community detection method designed by us is also presented. Additionally, the performance of these methods on the datasets with and without ground truth is compared. In conclusion, this comprehensive review provides a deep understanding of community detection in graphs. 
\end{abstract}

\vspace{5mm}


% \vspace{-12pt}
% \epigraph{``The greatest obstacle to discovery is not ignorance; it is the illusion of knowledge.''}{\textit{- Daniel J. Boorstin}}
% \vspace{-12pt}

\section{Introduction}

\subsection{Motivation}

Real-world systems, such as social networks, biological networks, and technological networks, exhibit complex structures and interactions. These systems can often be represented as graphs, where vertices represent individuals and edges represent interactions between them. Community detection, a specialized field of study, aims to decipher the complex structures and interactions within these graphs by clustering nodes based on the information provided by the graph. This research seeks to contribute to the field of community detection by conducting a comprehensive review of various methods and approaches. It also aims to bridge interdisciplinary perspectives by exploring community detection methods developed by statistical physicists and their applications in diverse domains. By understanding the community structure within these systems, valuable insights can be gained into their functioning and behavior, ultimately advancing our understanding of complex systems.

\subsection{Contribution} 

% This comprehensive review offers a thorough examination of community detection in graphs, elucidating four primary categories of mainstream frameworks. Within each framework, various detailed algorithms are explored, enriching the understanding of community detection methodologies. By delving into these algorithms, valuable insights into the intricate functioning and behavioral patterns of real-world systems are unveiled through the lens of their community structures. This elucidation not only facilitates a deeper comprehension of the mechanisms driving complex systems but also underscores the significance of community detection in analyzing and interpreting diverse datasets. Through this comprehensive approach, the review contributes to advancing our understanding of network dynamics and their implications across various domains.

This comprehensive review examines community detection in graphs, categorizing mainstream frameworks into four categories. It explores detailed methods within each framework, providing valuable insights into the community structures of real-world systems. By conducting comparative analysis using methods from each framework, the review further enhances our understanding of network dynamics and their implications across domains, contributing to advancements in complex system analysis.

Additionally, the review introduces the Revised Medoid-Shift (RMS) method as an innovative solution for the community detection problem. The RMS method combines the principles of Medoid-Shift and K-nearest neighbors (KNN) to improve community detection performance. By doing so, it overcomes the limitations of traditional machine learning methods, such as Mean-Shift, which struggle with non-Euclidean data structures like graphs. Through the integration of the RMS method, this research expands the toolkit available for network analysis and fosters deeper insights into the intricate structures and behaviors inherent in complex systems.

\subsection{Road Map}

This section outlines the structure and roadmap of the paper, aimed at providing a comprehensive understanding of the topic.

Section \ref{Research Background} delves into the research background, providing a detailed overview of the context and motivation behind this study. By presenting a thorough understanding of the research background, we lay the foundation for the subsequent sections.

Section \ref{method} explores the state-of-the-art methods in the field through an extensive review of the existing literature. We discuss various approaches and techniques employed by researchers, highlighting specific methods that are referenced and relevant to our study. Additionally, we introduce the new algorithm we constructed in detail. By examining these methods, we provide a comprehensive overview of the existing landscape and identify the strengths and limitations of different approaches.

Sections \ref{result} present the research findings and engage in in-depth discussions. We present the empirical results obtained through our research methodology, analyze and interpret the data, drawing meaningful insights and observations. We critically evaluate the implications of the findings, address potential limitations, and explore future research directions. By engaging in thorough discussions, we provide a comprehensive analysis of the research outcomes and contribute to the broader scientific discourse on the topic.

Section \ref{conclusion} concludes the paper by synthesizing the key findings, highlighting the contributions of our study, and discussing implications for future research and practical applications. By providing a conclusive summary, we aim to leave readers with a clear understanding of the research outcomes and inspire further exploration in the field.

Through this roadmap, we ensure a logical flow of ideas, enabling readers to engage with the research background, state-of-the-art methods, research findings, and final conclusions.

\section{Research Background of Community Detection}

\label{Research Background}

Community detection in graphs \citep{Fortunato2010,Sarkar2011,Bruna2017,Shchur2019,Lancichinetti2009} has garnered significant attention in various domains, including social network analysis \citep{Freeman2004,Knoke2019,Scott2011}, biology \citep{Krzakala2013,Traag2019,Sabater2007}, and information retrieval \citep{Kicsi2019,Das2020,Yan2012,Papadopoulos2012,DAmore2004,deBerardinis2020,Yang2009,Dridi2014,lai2024sharedprivateinformationlearning}. The pursuit of identifying cohesive groups of nodes characterized by strong intra-group connections and relatively weaker inter-group connections serves as a fundamental objective in network analysis \citep{Gleyze2013}. By discerning the community structure embedded within complex networks, researchers gain invaluable insights into the underlying organization, dynamics, and functionality of these intricate systems. Beyond mere delineation of clusters, understanding the intricate interplay among these communities unveils the nuanced behaviors and emergent properties inherent in complex systems. This deeper comprehension is pivotal not only for theoretical understanding but also for practical applications across various domains, including social networks, biological systems, and information networks. Moreover, it fosters the development of robust methodologies and tools, such as the RMS method, that is adept at navigating the complexities of community detection tasks, thereby facilitating more accurate and insightful analyses. As such, the endeavor to uncover and interpret community structures within complex networks stands as a cornerstone in the broader pursuit of understanding and harnessing the power of interconnected systems in diverse fields.

In recent years, numerous studies have focused on developing effective methods for community detection \citep{Papadopoulos2012duplicate,Barber2007,Yang2016,Yang2013,Leskovec2010,Christensen2023,Rostami2023,Gandomi2015,shen2023advancing,shen2023git,shen2023pbsl,shen2023triplet,shen2021exploring}. Early work in the field, such as the groundbreaking research by Girvan and Newman \citep{Girvan2002} on edge betweenness, laid the foundation for subsequent advancements. Since then, a plethora of methods and techniques have been proposed, each with its own strengths, limitations, and underlying principles.

One prevalent approach in community detection is based on modularity optimization, introduced by \citep{Girvan2002nest}. Modularity measures the quality of a partition by comparing the density of edges within communities to that of a null model \citep{Chen2014}. The optimization process aims to maximize the modularity score \citep{Rosvall2007}, identifying the partition that exhibits the highest level of community structure. Notable variations and extensions of modularity-based methods include the Louvain method \citep{Que2015}, the Infomap method \citep{Zeng2018}, and the Kernighan-Lin method \citep{Dutt1993}.

Another prominent approach in community detection involves spectral clustering techniques. Spectral methods leverage the eigenvectors and eigenvalues of the graph Laplacian matrix to identify communities \citep{Berahmand2022,Gulikers2017,Berahmand2021,Chen2017,Wai2019}. Notable spectral clustering methods include the Normalized Cut method \citep{Xu2009} and the RatioCut method \citep{Wei1989}. These methods have demonstrated effectiveness in uncovering communities in various domains, particularly when the underlying network exhibits clear spectral properties.

Furthermore, researchers have explored probabilistic models for community detection, such as the Stochastic Block Model (SBM) \citep{Abbe2017,Abbe2015,Lee2019}. SBM assumes that nodes within the same community have a higher probability of being connected compared to nodes in different communities. Inference techniques, such as maximum likelihood estimation and Bayesian inference, have been employed to estimate the parameters of the SBM and detect communities accurately \citep{Jin2015,Leung2009,Chen2018,Wong2020,Xie2013}.

With the widespread usage of deep learning, there has been a growing interest in incorporating deep learning techniques into community detection. Researchers have explored the use of deep learning models \citep{Bhatt2021,Goodfellow2016,Mosavi2019,Janiesch2021,Mathew2021,Wang2017}, Graph Neural Networks (GNNs) \citep{Zhou2020,Wu2020,Xu2018,Dwivedi2020}, and reinforcement learning \citep{costa2021towards,tu2018unified,alipour2022multiagent,xu2020community,zhao2012topic,zhang2020seal} for community detection tasks. These approaches leverage the expressive power of neural networks to capture complex patterns and dependencies within networks, enabling more accurate and robust community detection \citep{Li2023,deSapienzaLuna2023,ZhangZ2021}.

While significant progress has been made in community detection, several challenges persist. These include the detection of overlapping communities, the resolution limit problem, scalability issues for large-scale networks, and the robustness of methods to noise and perturbations \citep{Mandala2013,ZhangP2014,Akbar2020,Grover2016}. Addressing these challenges is crucial for advancing the field and developing more effective community detection methods.

Modularity optimization \citep{ZhangXS2009}, spectral clustering \citep{VanGennip2013,Li2018,Hu2020,VanLierde2019}, probabilistic models, and machine learning techniques \citep{Liu2020} all contribute to the advancement of this field. By leveraging these methods and addressing the existing challenges, researchers aim to gain a deeper understanding of complex networks and extract meaningful insights from various domains.

    \section{Overview of The Community Detection Methodology and Our Methodology} 

    \label{method}

    \begin{figure}[ht]
      \centering
      \includegraphics[width=1.0\linewidth]{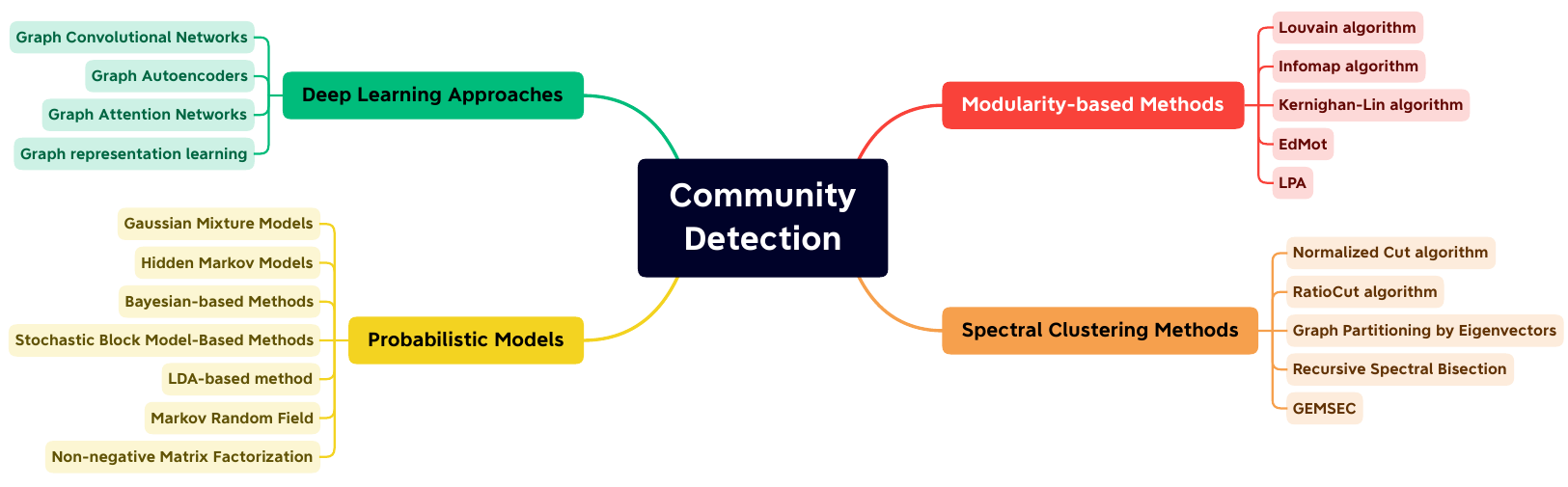}
      \caption{Overview of the community detection methodology}
      \label{fig:2}
  \end{figure}

    This section of this study delves into a comprehensive exploration of the current state-of-the-art methods in the field of community detection. Through an extensive review of the existing literature, we aim to shed light on the diverse range of approaches and techniques employed by researchers. This section serves as a platform to discuss and analyze various methods that have been referenced and deemed relevant to our study. Furthermore, we present a detailed introduction to the novel method developed as part of this research. By thoroughly examining these methods, our objective is to provide readers with a comprehensive overview of the existing landscape, enabling a better understanding of the strengths and limitations associated with different community detection approaches.
    
There are several types of methods and models for community detection in graphs. Here are four popular types along with examples of models and methods for each type:
    
\subsection{Modularity-based Methods}

Modularity-based methods are widely used in community detection and are based on the concept of modularity \citep{jin2020mod,Yang2016,Waltman2013}. Modularity measures the quality of a partition by comparing the density of edges within communities to that of a null model. The goal is to maximize the modularity score, which indicates the presence of a strong community structure within the network.

Modularity-based methods, as one of the prominent approaches in community detection, focus on identifying the partition of a network that maximizes the modularity score. This score serves as a metric to quantify the degree of segregation or clustering within the network, thereby revealing the presence of distinct and cohesive communities. By optimizing the modularity function, these methods aim to uncover meaningful structures inherent in the network topology, characterized by dense connections within communities and sparse connections between them. However, it's crucial to note that while modularity-based methods have demonstrated efficacy in identifying communities in various types of networks, they are not without limitations. Challenges such as resolution limit and sensitivity to network size and density have been observed, underscoring the need for nuanced approaches and careful interpretation of results. Despite these challenges, modularity-based methods remain valuable tools in the arsenal of community detection techniques, offering insights into the underlying organizational principles of complex systems across diverse domains, ranging from social networks to biological networks and beyond.

Modularity-based methods offer a diverse array of techniques for community detection, each presenting unique advantages and constraints. These methodologies serve as pillars upon which researchers continually build, refine and expand their capabilities to enhance both the precision and scalability of community detection methods. As the field advances, efforts are directed towards exploring Innovative variations and extensions of these methods. By embracing this iterative process of innovation and refinement, the quest for more accurate and scalable community detection solutions remains at the forefront of research endeavors.

\subsubsection{Louvain method}

\textbf{The Louvain method} \citep{Blondel2008} stands out as a widely acclaimed and efficient approach for modularity optimization within community detection. Its methodology revolves around a two-phase process, each meticulously designed to enhance the identification of meaningful community structures within networks. In the initial phase, the method meticulously delves into local optimization, meticulously refining the modularity within small communities. Following this, in the second phase, it seamlessly aggregates these optimized communities into larger, cohesive units, thereby providing a holistic view of the network's organizational landscape. This iterative process continues until reaching a convergence point, where no further enhancement in modularity is feasible. Noteworthy is the Louvain method's remarkable balance between speed and accuracy, rendering it not only a swift solution but also a precise one, effectively capturing intricate community structures within diverse networks. Its widespread adoption and consistently high performance underscore its significance as a cornerstone in the realm of community detection methodologies.

% 我后面都会用这个注释去标注这一块的引用方式
% (PMCDM: Privacy-preserving multiresolution community detection in multiplex networks)
\textbf{PMCDM: Privacy-Preserving Multiresolution Community Detection in Multiplex Networks} \citep{shao2022pmcdm} offers an innovative solution for uncovering community structures across various resolution levels in multiplex networks, with a strong focus on safeguarding the privacy of link and weight information. By utilizing differential privacy and homomorphic encryption techniques, PMCDM secures sensitive network data while employing a novel encrypted multi-greedy method, DH-Louvain, to optimize weighted modularity density. This approach ensures accurate identification of community structures without compromising privacy. Extensive experiments on both synthetic LFR benchmark networks and real-world datasets validate PMCDM's effectiveness in delivering precise community detection results, while simultaneously achieving high levels of privacy preservation. PMCDM represents a significant advancement in the analysis of complex systems within IoT environments, addressing both the challenges of multiresolution community detection and the critical need for privacy protection.

%(Scalable Community Detection with the Louvain Algorithm)
\textbf{PLA: Parallel Louvain Method} \citep{que2015scalable} is an advancement of the traditional Louvain method, designed for distributed memory systems to handle graphs with billions of edges efficiently. The method employs a novel graph mapping and data representation technique, supported by an efficient communication runtime, to enable fine-grained parallelism on supercomputers. It addresses the scalability challenge in community detection, making it possible to analyze massive graphs rapidly. The implementation can parallelize graphs with up to 138 billion edges across thousands of nodes, showcasing exceptional scalability. The PLA's ability to maintain or slightly improve the convergence properties, modularity, and quality of detected communities, compared to the original Louvain method, marks a significant achievement. This enhancement in community detection methodology facilitates the analysis of large-scale networks in various domains, from social networks to biology, with improved accuracy and efficiency.

% Distributed Louvain Algorithm for Graph Community Detection)
\textbf{DLA: Distributed Louvain Method} \citep{ghosh2018distributed} enhances community detection across distributed systems by leveraging the original Louvain method, tailored for scalability and efficiency in handling vast networks. Focused on modularity optimization, DLA distributes the graph across multiple nodes, incorporating heuristics to minimize communication overhead and synchronize community state updates efficiently. Tested on diverse real-world networks, DLA achieves remarkable scalability, processing graphs with billions of edges across thousands of processors without sacrificing the quality of community detection. This scalability not only underscores DLA's superior performance over existing methods but also its potential to facilitate detailed network analysis in various domains, from social networks to biological systems, thereby advancing the understanding of complex network structures through efficient community detection.

% The RNLA (Faster unfolding of communities: Speeding up the Louvain algorithm ) 
\textbf{RNLA: Random Neighbor Louvain Algorithm} \citep{traag2015faster} offers a groundbreaking improvement over the original Louvain method for community detection in complex networks. By selecting a random neighbor's community for a node's potential move, RNLA significantly reduces runtime complexity and achieves faster community detection without substantially compromising the quality of the results. This method not only simplifies computational efforts but also capitalizes on the structure of networks, where a random neighbor's community is often a suitable choice for a node's reallocation, enhancing the method's efficiency. The introduction of RNLA represents a significant step forward in the field of network analysis, offering a scalable, effective solution for identifying communities in vast and intricate networks, with broad applicability across various types of networks and quality measures.

%The LDA (From Louvain to Leiden: guaranteeing well-connected communities)
\textbf{LDA: Leiden Algorithm} \citep{traag2019louvain} innovatively enhances community detection in networks by addressing the shortcomings of the Louvain Algorithm, ensuring the creation of well-connected communities. By integrating smart local moves, fast local move approaches, and a refined aggregation process, LDA significantly improves both the speed and quality of community detection across diverse networks. This method guarantees that all identified communities are not only connected but optimally partitioned, through an iterative refinement process that converges to a partition where every subset of each community is locally optimally assigned. Its application to various benchmark and real-world networks demonstrates LDA's superior performance, making it an indispensable tool for the analysis of complex network structures.

%  (Parallelizing louvain algorithm: Distributed memory challenges)
\textbf{PLM-D: Parallel Louvain Method in Distributed Memory
The PLM-D} \citep{sattar2018parallelizing} innovatively parallelizes the Louvain method for enhanced community detection in vast networks, achieving notable performance gains. By leveraging both shared- and distributed-memory architectures, it attains up to a four-fold speedup in shared contexts, limited by physical core counts, and extends scalability in distributed settings via MPI (Message Passing Interface), surpassing previous caps of 16 processors. Key challenges identified include significant communication overhead, which restricts further scalability. PLM-D's contributions lie in its approach to distributed parallelization, providing insights into overcoming scalability barriers and suggesting future directions for reducing communication overhead and optimizing load distribution, thereby paving the way for scalable community detection on high-performance platforms.

% The RSNL (Self-adaptive Louvain algorithm: Fast and stable community detection algorithm based on the principle of small probability event)
\textbf{RSNL: Random Self-adaptive Neighbors Louvain method} \citep{zhang2018self} introduces a significant enhancement to community detection in complex networks by incorporating the principle of small probability events to optimize the selection of neighbors during the community detection process. This self-adaptive mechanism allows the method to maintain the speed and effectiveness of the original Louvain method while improving the accuracy and stability of the detected community structures. By comparing the RSNL's performance with the original Louvain method and the Random Neighbor Louvain method across both synthetic and real-world datasets, it is evident that RSNL achieves comparable or better community detection quality with increased computational efficiency. Moreover, the introduction of the equivalent computing time as a new metric demonstrates RSNL's superiority in achieving high-quality community partitions with fewer computational resources. This advancement not only enhances our understanding of complex network structures but also extends the applicability of community detection methods to larger and more complex datasets.

% The FKCD algorithm (Generalized Louvain method for community detection in large networks)
\textbf{FKCD: Fast K-path Community Detection} \citep{de2011generalized} enhances community detection in large networks by leveraging a novel edge centrality measure based on K-paths, efficiently computed in near-linear time. This method significantly advances the state-of-the-art by incorporating both global and local network information, facilitating the discovery of community structures through modularity optimization. Unlike traditional approaches that rely solely on local network features, FKCD's dual-focus strategy enables the identification of communities with higher precision and reliability, particularly in complex social networks. Experimentation on both synthetic and real-world datasets demonstrates FKCD's ability to outperform existing methods, offering a scalable solution for analyzing large networks. This breakthrough not only improves community detection accuracy but also extends its applicability to unweighted networks, positioning FKCD as a versatile tool in the field of network analysis.

% (Improving the Louvain algorithm for community detection with modularity maximization)
\textbf{Louvain+: Improved Louvain Algorithm for Community Detection with Modularity Maximization} \citep{gach2014improving} introduces an innovative enhancement to the renowned Louvain method for community detection by integrating a uncoarsening phase into the traditional multi-level modularity optimization framework. This extension transitions the Louvain method from a partial to a full multi-level method, incorporating both coarsening and uncoarsening phases to refine community detection results significantly. The core advancement lies in the uncoarsening-reﬁnement phase, which meticulously revisits and optimizes each community structure generated during the coarsening phase, ensuring a thorough exploration of the solution space for maximal modularity. Empirical evaluations on a diverse set of complex networks demonstrate that Louvain+ consistently outperforms the original Louvain method in terms of modularity, with only a modest increase in computational effort. This achievement underscores the efficacy of incorporating global optimization strategies in community detection, marking a significant step forward in the pursuit of discovering accurate and cohesive community structures in large networks.

%The C-Blondel algorithm (C-blondel: an efficient louvain-based dynamic community detection algorithm)
\textbf{C-Blondel: Compressed-Graph Based Louvain Dynamic Community Detection} \citep{seifikar2020c}  presents a novel enhancement in dynamic community detection by utilizing a compressed-graph approach based on previous network states, facilitating rapid and efficient identification of evolving communities within dynamic networks. By integrating the Louvain method's efficiency with the innovative concept of compressed graphs, where supernodes symbolize previously detected communities, C-Blondel significantly reduces computational complexity. This methodology not only expedites the community detection process through reduced graph size but also ensures high modularity, demonstrating its effectiveness through extensive testing on real-world datasets like Enron Email, Cit-HepTh, and Facebook. Through strategic handling of network changes, particularly by focusing on the removal of 'destructive nodes,' C-Blondel adeptly updates community structures, thereby offering a scalable and robust solution for tracking community evolution in large-scale dynamic networks. The method's success in maintaining comparable or superior modularity while substantially lowering execution times, as compared to its predecessors, underscores its potential as a valuable tool for contemporary network analysis.

% The LLPA algorithm (Improving Louvain algorithm for community detection)
\textbf{Louvain-LPA: Louvain-Label Propagation Algorithm Hybrid}
 \citep{hu2016improving} proposes an innovative hybrid approach to community detection in complex networks, seamlessly blending the speed of the Label Propagation Algorithm (LPA) with the modularity optimization of the Louvain method. By partitioning the network into two subsets based on node degree, LLPA applies the Louvain method to key nodes for enhanced quality and the LPA to edge nodes for reduced execution time, addressing the Louvain method's challenge of high time complexity in large-scale networks. This method not only preserves the accuracy of community detection but also significantly diminishes the method's time complexity. Experimental results on various real-world datasets demonstrate LLPA's capability to maintain or improve upon the Louvain method's modularity while achieving a notable reduction in execution time, making it an efficient and effective solution for community detection in evolving networks.

\subsubsection{Infomap method}

\textbf{The Infomap method} \citep{Rosvall2008}, grounded in the principles of information theory, offers a unique perspective on community detection by framing it as an optimization challenge aimed at minimizing the description length of a random walker's journey across the network. Unlike traditional approaches, Infomap delves into the intricate web of connections within a network, meticulously balancing the compression of trajectory within communities and the elucidation of transitions between them. By treating communities as efficient compression units and transitions as channels of information exchange, Infomap elegantly uncovers the underlying organizational structure of diverse networks. Its prowess lies in its ability to navigate through the network's complexity, efficiently identifying cohesive communities while accounting for the fluidity of interactions between them. Through rigorous optimization of its cost function, Infomap has consistently demonstrated its efficacy across a spectrum of network types, reaffirming its status as a versatile and reliable tool in the realm of community detection methodologies.

% The DIMA algorithm (A distributed infomap algorithm for scalable and high-quality community detection)
\textbf{DIMA: Distributed Infomap Algorithm} \citep{zeng2018distributed} innovates in the field of community detection within large networks by harnessing a distributed computing approach, effectively addressing the scalability challenge posed by the traditional Infomap method. Through a novel implementation of delegate partitioning and distribution methodologies, DIMA meticulously orchestrates the computation and communication within a distributed environment, achieving convergence of the clustering method while preserving the quality of community detection. By employing MPI for its framework, DIMA demonstrates exceptional adaptability to massively distributed systems. Experimental evaluations validate not only the method's correctness but also its superior scalability and efficiency when applied to extensive real-world datasets, showcasing remarkable performance improvements over previous implementations. DIMA sets a new benchmark for large-scale network analysis by delivering a scalable, accurate, and efficient solution for community detection, highlighting its potential to facilitate deeper insights into the complex structures of vast networks.

% The Infomap-SA algorithm (A Novel Algorithm Infomap-SA of Detecting Communities in Complex Networks)
\textbf{Infomap-SA: Infomap Simulated Annealing} \citep{hu2015novel} introduces a novel approach to community detection in complex networks by integrating the strengths of the Infomap method with the Simulated Annealing (SA) optimization technique. It leverages the optimization of modularity function, aiming to enhance the precision and efficiency of identifying community structures. This hybrid method uniquely encodes communities and nodes, employing a compression strategy to minimize coding length, thus effectively capturing the most cohesive community distribution. Experimental validation on real-world and computer-generated LFR-benchmark networks demonstrates that Infomap-SA not only achieves higher modularity and density but also exhibits reduced computational complexity compared to the traditional Infomap method. The infusion of SA allows Infomap-SA to navigate the solution space more effectively, overcoming limitations associated with poor global search capabilities and parameter sensitivity inherent in SA alone. As a result, Infomap-SA emerges as a more suitable candidate for large-scale network community detection, marking a significant advancement in the field of network analysis.

% InfoFlow (InfoFlow: A Distributed Algorithm to Detect Communities According to the Map Equation)
\textbf{InfoFlow: Distributed Algorithm for Community Detection via Map Equation} \citep{fung2019infoflow} represents a groundbreaking evolution in community detection methods, integrating distributed computing advancements with the foundational principles of the map equation. This method effectively addresses the challenges posed by large-scale networks, showcasing a shift from the linear complexity of its predecessor, InfoMap, to a logarithmic time complexity framework. By employing a novel approach that allows for parallel processing of multiple module mergers within a single iteration, InfoFlow significantly accelerates the community detection process without compromising the accuracy of the resulting partitions. The method's efficacy is demonstrated through benchmarks on graphs encompassing millions of nodes and hundreds of millions of edges, where it not only maintains high community detection accuracy but also showcases a dramatic reduction in computation time compared to traditional methods. This development not only marks a significant advancement in handling large network datasets but also opens new avenues for further research into distributed computing methods for network analysis.

\subsubsection{Kernighan-Lin method}

\textbf{The Kernighan-Lin method} \citep{Kernighan1970} stands as one of the pioneering modularity-based techniques in the realm of community detection. Its methodology relies on a pragmatic, greedy strategy aimed at enhancing modularity through iterative node reassignments across communities. Commencing with an initial partition, the method meticulously assesses the modularity gain associated with node exchanges between communities. This iterative process persists until reaching a point where no further enhancement in modularity can be discerned. While the Kernighan-Lin method may not invariably converge to the globally optimal solution, its computational efficiency renders it a pragmatic choice for community detection tasks. Its simplicity and speed make it particularly appealing for analyzing large-scale networks where exhaustive search strategies might prove prohibitively costly. Thus, while not without its limitations, the Kernighan-Lin method remains a valuable tool in the community detection toolkit, offering a balance between computational tractability and meaningful insights into network structure.

% The Combo algorithm (General optimization technique for high-quality community detection in complex networks)
\textbf{Combo: General Optimization Technique for High-Quality Community Detection} \citep{sobolevsky2014general} proposes a versatile and efficient optimization strategy for community detection across various complex networks, addressing the core challenge of optimizing different objective functions, notably modularity and description code length. This innovative approach combines merging, splitting, and node-moving strategies within communities to significantly surpass existing methods in optimizing modularity, evidenced by superior final scores on both synthetic and real-world networks. Additionally, when optimizing for description code length, Combo achieves comparable results to the Infomap method, suggesting its robust applicability across different objective functions. The method demonstrates an optimal balance between execution time and the quality of community partitioning, making it suitable for networks up to 30,000 nodes. The significant improvement in modularity scores obtained by Combo underscores the importance of precision in community detection, where even minor enhancements can lead to substantial differences in detected community structures. This advance not only sets a new benchmark in community detection methods but also opens avenues for exploring the optimization of other objective functions within complex network analyses.

% The MFTS algorithm (Improved community structure detection using a modified fine-tuning strategy)
\textbf{MFTS: Modified Fine-Tuning Strategy for Community Detection} \citep{sun2009improved} enhances the accuracy of community detection in complex networks by introducing an additional fine-tuning step to the conventional recursive bisection process, aimed at maximizing modularity without increasing the computational complexity order. This strategy addresses the inherent bias in size distribution of detected communities by allowing nodes to move across existing communities after the standard bisection and fine-tuning steps. By applying this modified strategy to the leading eigenvalue method, MFTS removes the undue constraints that limit the effectiveness of traditional methods, as demonstrated through significantly improved modularity scores on both synthetic and real-world networks. The inclusion of a final-tuning step, which considers the movement of each node to any community, results in a more accurate and unbiased detection of community structures, making MFTS competitive with, and in some cases superior to, the best existing community detection methods. This methodological enhancement underscores the importance of flexibility in community assignment for achieving high-quality network partitioning, establishing MFTS as a valuable tool for the analysis of complex network structures.

%The NMF algorithm (Efficient Community Detection in Large Scale Networks)
\textbf{NMF: Newman's Modularity Framework with Fine-Tuning} \citep{vieira2013efficient} offers a sophisticated approach to community detection in large-scale networks by efficiently implementing Newman's spectral method with an enhanced Fine-Tuning (FT) stage. This advancement significantly reduces execution time while maintaining the integrity of division quality, thanks to strategic data structure and algorithmic optimizations. By refining the FT process to limit node movements and leveraging sparse matrix representations, NMF achieves remarkable performance improvements. Experimental validations on networks ranging from small social systems to extensive web matrices demonstrate NMF's capability to handle over a million nodes efficiently. The incorporation of FT variations, particularly limiting node adjustments to 10-20\% of the network, effectively balances computational efficiency with community detection fidelity. NMF not only excels in scaling to real-world network sizes but also in providing a competitive alternative to existing methods, making it an essential tool for contemporary network analysis endeavors.

\subsubsection{EdMot}

\textbf{The Edge Enhancement approach for Motif-aware community detection (EdMot)} \citep{Li2019} represents a pioneering solution tailored to overcome the limitations inherent in conventional community detection methods. Existing approaches often concentrate on lower-order structures, such as individual nodes and edges, neglecting the nuanced characteristics of higher-order features encapsulated within small dense subgraph patterns known as motifs. Higher-order methods typically operate under the assumption of a connected hypergraph, a notion that falters when applied to real-world networks. In reality, hypergraphs may devolve into fragmented entities, comprising numerous connected components and isolated nodes, contrary to the original connected network. This fragmentation poses a formidable challenge for existing higher-order methods, as they struggle to unite nodes that lack connections in the hypergraph, even if they share a community affiliation. To combat this fragmentation dilemma, EdMot introduces an innovative edge enhancement approach.

\subsubsection{LPA}

% The LPA-MNI algorithm (LPA-MNI: an improved label propagation algorithm based on modularity and node importance for community detection)
\textbf{LPA-MNI: Label Propagation Algorithm Based on Modularity and Node Importance} \citep{li2021lpa} extends the conventional label propagation \citep{zhu2002} approach by integrating modularity optimization and node importance to tackle the inherent issues of randomness and instability in community detection. Specifically, it commences with identifying initial communities through modularity maximization, subsequently assigning unique labels to nodes within these initial communities, thereby reducing the method's susceptibility to randomness. The node importance, determined by degree centrality, dictates the order of label updates, ensuring a systematic and stable label propagation process. Extensive testing on a diverse array of both real-world and synthetic networks demonstrates LPA-MNI's enhanced accuracy, robustness, and efficiency in detecting community structures. Its ability to produce higher modularity and more consistent community numbers, as compared to six other leading methods, confirms its effectiveness. Furthermore, LPA-MNI's introduction of a deterministic and accurate label update mechanism significantly improves upon the traditional label propagation method's stability, presenting a formidable solution for community detection in complex networks.

% The CLPA (Detecting network communities by propagating labels under constraints)
\textbf{CLPA: Constrained Label Propagation Algorithm} \citep{barber2009detecting} advances community detection in networks by reformulating the label propagation algorithm into an optimization problem, aiming to address its inherent drawbacks. This adaptation involves the introduction of a modified objective function that incorporates constraints to penalize less desirable solutions, such as the trivial solution where all vertices share the same label. By proposing several constraint variants, including a modularity-maximizing variant and adaptations for bipartite networks, CLPA showcases a diverse approach to refining community structures. Through comparative analysis on both synthetic and real-world networks, CLPA not only demonstrates its ability to evade the pitfalls of standard LPA but also highlights its versatility in handling different network types. This methodology enables the detection of more meaningful communities by effectively leveraging constraints within the optimization process, thus offering a nuanced and flexible tool for network analysis.

\textbf{Near linear time algorithm to detect community structures in large-scale networks(label propagation )} \citep{Raghavan2007} under scrutiny in this study offers a straightforward and effective avenue for community detection within real-world networks. Diverging from existing methodologies that either necessitate prior knowledge of community structures or entail computationally intensive procedures, this method hinges solely on the inherent structure of the network to guide its operations. Its iterative nature begins by assigning each node a distinct label. Subsequently, at each iteration, nodes update their labels based on the prevailing label among their neighbors. This iterative label exchange process facilitates the emergence of densely interconnected clusters of nodes, gradually coalescing around a consensus label, thereby delineating distinct communities within the network. This simplistic yet powerful approach underscores the method's efficacy in efficiently discerning community structures without the need for additional information or imposing computational burdens. Moreover, its reliance on local interactions within the network fosters scalability, rendering it well-suited for analysis in large-scale network environments.

% LPAm+ (Advanced modularity-specialized label propagation algorithm for detecting communities in networks)
\textbf{LPAm+: Advanced Modularity-Specialized Label Propagation Algorithm} \citep{liu2010advanced} advances community detection by optimizing a modularity-specialized Label Propagation Algorithm (LPAm) with a Multi-step Greedy Agglomerative strategy (MGA) to escape local modularity maxima. By iteratively applying LPAm to refine community labels and employing MGA to merge communities that improve modularity most, LPAm+ successfully identifies higher-quality community structures in networks. This method addresses LPAm's tendency to converge to sub-optimal partitions by enabling it to navigate out of local modularity peaks, resulting in a significant enhancement in the accuracy of detected communities. Experiments across various real-world networks demonstrate LPAm+'s ability to achieve superior modularity scores compared to LPAm alone, providing a balance between computational efficiency and the precision of community detection. This approach not only improves upon the modularity values reported by existing methods but also maintains a practical execution time, positioning LPAm+ as an effective solution for community detection in complex networks.

%The NSLPA (Community Detection Using a Neighborhood Strength Driven Label Propagation Algorithm)
\textbf{NSLPA: Neighborhood Strength Driven Label Propagation Algorithm} \citep{xie2011community} represents a significant advancement in community detection within networks by introducing a novel label propagation criterion that incorporates the concept of neighborhood strength. This approach significantly improves both the computational efficiency and the accuracy of community delineation compared to the original Label Propagation Algorithm. By optimizing update rules to prevent unnecessary iterations and incorporating a generalized update rule that takes into account varying strengths of connections within a node's neighborhood, NSLPA demonstrates a marked reduction in computational time, especially for large networks. Experimental validations on both synthetic and real-world networks exhibit the method's superior capability to discern high-quality community structures, highlighting its effectiveness in leveraging local network properties to enhance community detection. This methodology not only expedites the detection process but also yields more meaningful and coherent community partitions, underscoring its potential for extensive application in analyzing complex network structures.

% The LPA-HM algorithm (A label propagation algorithm for community detection on high-mixed networks)
\textbf{LPA-HM: Label Propagation Algorithm for High-Mixed Networks} \citep{wu2021label} presents an innovative strategy for community detection within highly mixed networks by utilizing the node influence and a novel label propagation mechanism. This method significantly improves upon the standard label propagation algorithm by initially preprocessing node labels based on the number of common neighbors, effectively reducing the randomness in label assignment. During the propagation phase, the method strategically assigns labels to nodes by considering both the prevalence of labels among a node's neighbors and the overall influence of those labels, thus ensuring a more stable and accurate community detection process. Furthermore, LPA-HM incorporates early stopping criteria based on modularity and run-to-run changes in community numbers to prevent over-propagation of labels, addressing a common limitation of LPA in high-mixed networks. Through comprehensive experimentation on both real-world and synthetic networks, LPA-HM demonstrates superior performance in detecting communities with high accuracy, particularly in networks with a mixing parameter greater than or equal to 0.5, where conventional methods tend to falter. This approach not only enhances the detection of nuanced community structures in complex networks but also maintains the efficiency and scalability inherent to LPA, marking a significant advancement in the field of network analysis.

\textbf{TNS-LPA} \citep{Xu2020TNS} is an improved label propagation algorithm for community detection in complex networks. It introduces a new two-level neighborhood similarity measure called TNS, which combines resource allocation and local path similarity. The method selects initial community centers based on minimum distance and local centrality index to ensure their importance and spatial distribution. It employs the new similarity measure and an optimization strategy for asynchronously updating labels according to node importance. Additionally, the method incorporates label influence based on the new similarity measure to enhance the accuracy of community division. Experimental results demonstrate that TNS-LPA outperforms existing methods in terms of modularity, normalized mutual information, and adjusted rand index.

%The FLPA (Large network community detection by fast label propagation)
\textbf{FLPA: Fast Label Propagation Algorithm} \citep{traag2023large} proposes a highly efficient variant of the traditional Label Propagation Algorithm tailored for large-scale network community detection. By introducing a dynamic queue mechanism for nodes whose neighborhoods have recently changed, FLPA significantly accelerates the process of label updating, focusing computational resources on the most impactful nodes. This targeted approach allows FLPA to achieve remarkable speed improvements—up to 700 times faster than the standard LPA—without compromising the quality of the detected communities. Each node in the network is guaranteed to retain the majority label within its immediate community, ensuring the integrity of the community structure is preserved. FLPA's ability to maintain high accuracy in community detection, coupled with its substantial reduction in computational time, marks a significant advancement in the field of network analysis, making it a preferred method for initial examinations of network structures, particularly in applications requiring rapid processing of extensive datasets.

\textbf{LDA-LPA: Latent Dirichlet Allocation with Label Propagation Algorithm}
%The LDA-LPA (LDA-LPA: A Community Detection Method Based on Topic Model)
 \citep{Wang2021LDA} innovates in community detection within complex networks by fusing the LDA topic model with the Label Propagation Algorithm, effectively utilizing only the structural information of networks without requiring node or edge attributes. By interpreting each community as a distinct topic and each network node as a document, the LDA model assigns nodes to communities based on topic distributions derived from neighborhood-driven document representations. This process is followed by a refinement phase using LPA, which leverages initial community assignments to enhance the precision of community detection. The dual-phase approach ensures a balance between the global structural insights of LDA and the local optimization capabilities of LPA, demonstrating superior performance over benchmark methods in both real-world and synthetic networks. This method's reliance solely on network topology for community detection broadens its applicability and introduces a novel graph embedding technique based on node neighborhoods, further streamlining the community detection process without the need for an extensive learning phase.

\textbf{DLPA: Directed Balanced Label Propagation Algorithm}
%The DLPA (A label propagation-based method for community detection in directed signed social networks) 
 \citep{hosseini2022label} introduces an innovative method for detecting communities in directed signed social networks, focusing on optimizing community detection by incorporating edge direction and signed relationships. By converting directed signed social networks into weighted undirected signed social networks, DLPA assigns weights based on edge direction and sign information. It extends the Label Propagation Algorithm by integrating these weights into the label propagation process, enhancing the method's ability to identify community structures accurately. The method's effectiveness is demonstrated through extensive experiments on both real-world and synthetic datasets, where DLPA outperforms existing approaches in terms of directed signed modularity, directed signed flow-based capacity, and frustration measures, highlighting its advantages in capturing the complex dynamics of directed signed social networks while acknowledging its computational complexity.

\textbf{NI-LPA: Node Importance-based Label Propagation Algorithm}
%The NI-LPA (Node Importance based Label Propagation Algorithm for overlapping community detection in networks)
 \citep{el2020node} for detecting overlapping communities in networks enhances the conventional Label Propagation Algorithm by integrating a novel measure of node importance, aiming at improving stability and accuracy. Unlike traditional LPA, which treats all nodes equally, NI-LPA assigns each node a unique importance value based on its degree and clustering coefficient. This approach addresses LPA's randomness by initializing nodes with a unique label and refining community detection through a specialized propagation and filtering process that leverages node properties. Experimental results on both artificial and real-world networks demonstrate NI-LPA's superior performance in identifying overlapping communities, showing significant improvements in efficiency and precision. By resolving LPA's instability and incorporating node importance, NI-LPA emerges as a robust and effective method for community detection in complex networks, offering insights into network structures with overlapping community memberships.

\textbf{MSH-LPA: Multilayer SH-Index-Based Label Propagation Algorithm}
%The MSH-LPA (A Feasible Community Detection Algorithm for Multilayer Networks)
 \citep{chen2020feasible} advances community detection in multilayer networks through an innovative integration of the SH-index and a novel multilayer network aggregation model. This approach enhances the traditional label propagation algorithm by incorporating node centrality and edge weights to mitigate its inherent instability. Specifically, the MSH-LPA method calculates the centrality of nodes using the SH-index and adjusts the weight of edges based on node similarity within a weighted single-layer network derived from the original multilayer structure. The method's efficacy is validated through comprehensive experiments on various datasets, demonstrating its superior ability to achieve higher modularity and more accurate community detection compared to existing methods. The MSH-LPA's introduction of weighted edges and centrality measures significantly improves upon the precision and stability of community detection in complex, multilayer networks, despite its computational complexity remaining close to linear time, thus presenting a balanced approach between computational efficiency and detection accuracy.

\subsection{Spectral Clustering Methods}

Spectral clustering methods are a powerful family of methods that exploit the spectral properties of graphs to detect communities. By leveraging the eigenvalues and eigenvectors of the graph Laplacian matrix, these methods can effectively partition nodes into distinct groups, making them particularly adept at identifying communities with complex topological structures. Due to their effectiveness, spectral clustering approaches have gained significant popularity in the field of community detection.
One key advantage of spectral clustering is its ability to handle various types of network structures. Whether the network is sparse, dense, or contains disconnected components, spectral clustering can adapt and provide meaningful community detection results. This adaptability stems from its reliance on the eigenvalues and eigenvectors, which capture important structural information about the network.
Moreover, spectral clustering methods are not limited to a specific type of similarity matrix. While the adjacency matrix and the normalized Laplacian matrix are common choices, spectral clustering can also be applied using other similarity measures or a combination of different matrices. This flexibility allows spectral clustering to be applied to a wide range of networks, from social networks to biological networks, effectively capturing their underlying community structure.

The general steps involved in spectral clustering are as follows:
First step, constructing the similarity matrix: The first step is to construct a similarity matrix that captures the relationships between nodes in the graph. Common choices for the similarity matrix include the adjacency matrix, the normalized Laplacian matrix, or a combination of both.

Second step, computing the graph laplacian: The graph Laplacian matrix is derived from the similarity matrix and plays a crucial role in spectral clustering. There are different types of graph Laplacians, such as the unnormalized Laplacian, the normalized Laplacian, and the symmetric normalized Laplacian. Each variant has its own characteristics and implications for community detection.

Third step, eigenvalue decomposition: The next step involves decomposing the graph Laplacian matrix to obtain its eigenvalues and corresponding eigenvectors. The number of eigenvalues and eigenvectors considered depends on the desired number of communities to be detected.

Fourth step, Clustering Eigenvectors: After obtaining the eigenvectors, clustering methods such as k-means or normalized cut are applied to group the eigenvectors into distinct communities. The specific clustering method used may vary depending on the spectral clustering variant.

% \subsubsection{Adjacency Matrix-based Spectral Clustering}

\subsubsection{Normalized Cut algorithm}

\textbf{Normalized Cut algorithm} \citep{Shi2000}stands out as a widely embraced spectral clustering method, recognized for its efficacy in community detection. At its core, this method endeavors to minimize the normalized cut value, serving as a quantitative measure for the dissimilarity between distinct communities within a graph. The optimization process is meticulously executed as the method strategically partitions the graph into communities, ensuring the minimization of the normalized cut objective. To delve into the intricacies of its functionality, the method engages in a sophisticated optimization procedure. This involves performing eigenvalue decomposition on the normalized Laplacian matrix, a crucial step that unveils the underlying spectral properties of the graph. Through this decomposition, the method extracts eigenvalues and eigenvectors, unraveling essential structural information about the network. With these extracted eigenvectors in hand, the Normalized Cut algorithm proceeds to employ a clustering method. This phase is integral in grouping nodes into cohesive communities based on the information gleaned from the eigenvalue decomposition. The meticulous orchestration of eigenvalue analysis and subsequent clustering ensures a robust and accurate identification of communities within the graph.

\subsubsection{RatioCut algorithm}
\textbf{RatioCut algorithm} \citep{Hagen1992}, much like its counterpart, the Normalized Cut algorithm, is designed to minimize a dissimilarity measure known as the ratio cut. This cut is essentially a quantification of the ratio between the number of edges connecting different communities and the total number of edges within each individual community. The primary objective of the RatioCut algorithm is to pinpoint and segregate distinct communities within a given graph by reducing this ratio to a minimum. To achieve this, the method leverages eigenvalue decomposition, a mathematical technique used to break down the graph's properties into fundamental components. Subsequently, a clustering method is applied to the decomposed eigenvalues, facilitating the partitioning of the graph into cohesive and well-defined communities. This combination of ratio cut optimization, eigenvalue decomposition, and clustering forms a robust approach for uncovering meaningful structures within complex networks.

\subsubsection{Graph Partitioning by Eigenvectors}
\textbf{Graph Partitioning by Eigenvectors} \citep{Pothen1990} is a sophisticated method employed in network analysis to uncover underlying communities within graphs. Central to its operation are the leading eigenvectors derived from the graph's Laplacian matrix, which encapsulates crucial structural information. Through meticulous examination of these eigenvectors, the method discerns intricate patterns and associations within the network. Its spectral bisection technique facilitates the hierarchical division of the graph, systematically segregating nodes into subsets based on the polarity of their eigenvector components. This iterative approach refines the partitioning iteratively until the desired number of communities emerges. The flexibility of this method lies in its ability to adapt to diverse datasets, enabling analysts to tailor the partitioning process to suit specific analytical requirements. As a result, the Graph Partitioning by Eigenvectors method stands as a powerful and versatile tool, invaluable across domains such as social network analysis, biological network modeling, and beyond. Its systematic approach offers insights into complex network structures, unlocking opportunities for deeper understanding and meaningful interpretation.

\textbf{The Social Network Community Detection Using Agglomerative Spectral Clustering method (ASC) } \citep{narantsatsralt2017} leverages spectral clustering enhanced by agglomeration based on conductance and edge weights to refine community detection in complex networks. This innovative approach improves upon traditional spectral clustering by utilizing a similarity function in the eigenvector space and edge weights for more precise node aggregation. Conductance serves as a critical termination criterion for agglomeration, ensuring densely connected clusters. Experiments on real-world networks, such as LiveJournal and Orkut, demonstrated the method's effectiveness, showing superior performance in detecting intricate community structures. Despite its success, the method faces challenges in computational efficiency and the precise determination of the optimal number of communities, indicating potential areas for further enhancement.

\subsubsection{Recursive Spectral Bisection}
\textbf{Recursive Spectral Bisection} 
 \citep{Spielman2007} represents an innovative extension of the Graph Partitioning by Eigenvectors method, enhancing its capabilities in uncovering communities within complex networks. Unlike its predecessor, which partitions the graph into disjoint subsets through spectral bisection, this method introduces a recursive bipartitioning strategy. By dividing the graph into smaller subgraphs and iteratively applying spectral bisection on each subgraph, the method achieves a multi-scale community detection approach. This recursive nature enables the method to capture communities at various granularities, accommodating the diverse structural intricacies present in real-world networks. Moreover, by adaptively adjusting the partitioning process based on the local structural characteristics of each subgraph, the method exhibits robustness and scalability across a wide range of network sizes and densities. Through its systematic and scalable approach, the Recursive Spectral Bisection method offers a powerful tool for network analysis in fields such as social sciences, biology, and computer science. Its ability to reveal communities at different scales provides researchers with a comprehensive understanding of network structures, facilitating insights into underlying patterns and dynamics critical for informed decision-making and problem-solving.

\textbf{GRSBM: Greedy Recursive Spectral Bisection for Modularity-Bound Hierarchical Divisive Community Detection} \citep{cardoso2024} introduces a sophisticated method for community detection in networks, employing a greedy recursive spectral bisection approach grounded in modularity optimization. Leveraging the principles of spectral graph theory, GRSBM commences with a network sparsification step to highlight community structures, followed by iterative application of spectral bisection, utilizing the eigenvector associated with the second smallest eigenvalue of the normalized Laplacian matrix. This process recursively divides the network into increasingly finer sub-communities, guided by modularity maximization at each step to ensure the relevance of the identified communities. Through comprehensive experimental evaluations across diverse real-world and synthetic networks, GRSBM exhibits superior performance in accurately discerning community structures, showcasing a significant improvement in both modularity and the adjusted Rand index over existing methods. This method's ability to adaptively determine the number of communities without prior specification and its robustness in handling complex network topologies underscore its advanced contribution to the field of network analysis.

\subsubsection{GEMSEC}
\textbf{GEMSEC: graph embedding with self clustering} \citep{Rozemberczki2019} represents a significant advancement in graph embedding techniques, seamlessly integrating node embedding learning with clustering tasks. Building upon prior research in sequence-based graph embedding, GEMSEC introduces innovative methodologies tailored to leverage known social network properties effectively. Central to its operation is the abstraction of nodes into a feature space, wherein vertex features are strategically optimized to minimize the negative log-likelihood of preserving sampled vertex neighborhoods. By doing so, GEMSEC ensures that the learned embeddings accurately capture the intricate local connectivity patterns inherent within the graph structure. Moreover, GEMSEC distinguishes itself by incorporating sophisticated machine learning regularization techniques, which harness prior knowledge regarding the network's structure and dynamics. Through these enhancements, GEMSEC not only enhances the efficiency of processing large graphs but also elevates the quality and interpretability of the resulting embeddings, thereby facilitating deeper insights into complex network phenomena.

\subsection{Probabilistic Models}

Probabilistic models stand as a versatile class of frameworks leveraging probability theory to encapsulate uncertainty, enabling the prediction and inference of outcomes based on available data. They furnish a structured approach to apprehending and quantifying uncertainties inherent in real-world phenomena, finding extensive application across disciplines such as machine learning, statistics, and artificial intelligence. These models prove indispensable for modeling intricate systems and accommodating incomplete or noisy datasets. Typically, probabilistic models comprise key elements including variables representing aspects of interest, parameters governing their behavior, probability distributions delineating the likelihood of different outcomes, and mechanisms for inference, facilitating the extraction of meaningful insights from the modeled data. Through the interplay of these components, probabilistic models offer a powerful framework for understanding and navigating uncertainty in diverse domains.

Variables, the first foundational component of probabilistic models, serve as the entities of interest within the model framework. These can manifest as either observed variables, representing the measurable data directly available for analysis, or latent variables, which remain hidden but are inferred from the observed data. The dynamic interplay between observed and latent variables allows for a comprehensive representation of complex systems, accommodating both tangible measurements and underlying, unobservable factors.
Parameters, the second critical element, play a pivotal role in defining the probabilistic relationships between variables within the model. These values essentially govern the behavior of the model, acting as the building blocks that shape the overall structure. Derived from the available data through estimation processes, parameters encapsulate the essential characteristics of the system under investigation. Intriguingly, parameters themselves can exhibit diverse properties; they may be fixed constants, providing stability to the model, or they can take on the form of random variables, introducing an additional layer of uncertainty into the model's architecture. This versatility allows probabilistic models to adapt to a wide array of scenarios, from deterministic relationships to those influenced by inherent randomness.

Probability distributions serve as mathematical models that elucidate the probability of various outcomes for variables or parameters within a given context. By delineating the likelihood of observing specific values, they offer crucial insights into the uncertainty inherent in various phenomena. These distributions encapsulate the probabilities associated with different outcomes, often conditioned on the values of other variables or parameters. Among the myriad of probability distributions, several stand out as fundamental in diverse fields. The Gaussian (normal) distribution, for instance, is ubiquitous due to its prevalence in natural phenomena and its mathematical tractability. The Bernoulli distribution is vital in modeling binary outcomes, making it indispensable in areas such as coin flipping or binary classification tasks. Meanwhile, the multinomial distribution extends this concept to scenarios involving multiple categorical outcomes, such as the roll of a dice or the outcomes of a survey with multiple response options. Moreover, the Dirichlet distribution emerges as a cornerstone in Bayesian statistics, particularly in modeling distributions over probability vectors, making it invaluable in tasks like topic modeling or Bayesian inference. These distributions collectively form the backbone of statistical analysis, empowering researchers and practitioners across various domains to quantify uncertainty and make informed decisions.

Inference, a pivotal stage in statistical analysis, encompasses the art and science of deriving meaningful insights from probabilistic models. At its core, inference endeavors to unravel the hidden facets of a system by estimating the values of latent variables or parameters based on observed data. This process serves as a bridge between raw information and actionable knowledge, facilitating decision-making and hypothesis testing across various domains. One of the fundamental techniques employed in inference is maximum likelihood estimation (MLE), which strives to find the parameter values that maximize the likelihood of observing the given data. While MLE provides a straightforward approach, Bayesian inference offers a more comprehensive framework by incorporating prior knowledge and updating beliefs in light of new evidence. By leveraging Bayesian principles, practitioners can not only estimate parameters but also quantify uncertainty and refine their models iteratively. Additionally, variational inference presents a versatile alternative, aiming to approximate complex posterior distributions through optimization-based methods. This approach strikes a balance between accuracy and computational efficiency, making it suitable for large-scale datasets and intricate models. As the cornerstone of statistical reasoning, inference empowers researchers and analysts to extract meaningful insights from data, guiding decision-making and fostering a deeper understanding of the underlying phenomena.

\subsubsection{Gaussian Mixture Models (GMM)}

\textbf{GMM} \citep{Dempster1977} stands as a powerful probabilistic framework designed to characterize data by expressing it as a blend of Gaussian distributions. This model posits that the observed data points arise from a complex interplay of multiple Gaussian distributions, each characterized by its unique mean and covariance. The elegance of GMM lies in its ability to capture the underlying structure of diverse datasets, accommodating variations and nuances that may not be adequately represented by a single Gaussian distribution.

At its essence, GMM serves as a flexible tool for modeling complex data patterns, offering a nuanced understanding of the underlying processes governing the observed phenomena. The fundamental idea is that the data is not generated by a single Gaussian distribution but rather by a mixture of these distributions, allowing for a more accurate representation of the inherent variability within the dataset.

To harness the potential of GMM, the model parameters, encompassing means, covariances, and mixture weights, need to be estimated. The Expectation-Maximization (EM) method emerges as the key driver in this process. EM operates iteratively, iteratively refining its estimates of the model parameters until convergence is achieved. The "Expectation" step involves estimating the probability that each data point belongs to each Gaussian component, while the "Maximization" step updates the parameters based on these probabilities. This cyclic process continues until the model converges to a configuration that maximizes the likelihood of the observed data under the GMM.

In practical terms, GMM and the EM methods find extensive application in various domains, ranging from image and speech processing to finance and biology. Their versatility makes them particularly adept at capturing complex patterns and extracting valuable insights from data, fostering a deeper understanding of the intricate structures that underlie diverse datasets. As an indispensable tool in the statistical toolkit, GMM with the EM method empowers analysts and researchers to unravel the hidden intricacies of data, providing a robust foundation for informed decision-making and hypothesis testing \citep{Moon1996}.

%(Community Detection in Multi-relational Social Networks)
\textbf{MutuRank algorithm} \citep{wu2013community} aims at detecting communities within multi-relational social networks. Unlike previous methods that treat relations independently, MutuRank acknowledges the mutual influence between relations and actors by transforming multi-relational networks into single-relational networks for more effective community detection. The core of MutuRank lies in its novel co-ranking framework, which determines the weights of various relation types and nodes through iterative updates, achieving equilibrium probability distributions that more accurately reflect the intrinsic status of relations and objects. Following this transformation, the Gaussian Mixture Model with Neighbor Knowledge (GMM-NK) is applied, utilizing local consistency principles to enhance spectral clustering's performance in identifying overlapping communities. Experimental results on both synthetic and real-world datasets, such as DBLP, demonstrate MutuRank combined with GMM-NK's effectiveness over existing methods, showcasing its ability to reveal clearer community structures by integrating multi-relational data and detecting overlapping communities through a probabilistic approach.

% (Learning Community Embedding with  Community Detection and Node Embedding on Graphs)
\textbf{ComE} \citep{Cavallari2017} introduces a framework for learning community embedding by integrating community detection and node embedding on graphs. This approach is predicated on the understanding that community embedding benefits from a symbiotic relationship with community detection and node classification, thereby forming a closed loop that enhances the effectiveness of graph visualization, community detection, and node classification tasks. Specifically, the ComE framework represents community embedding as a multivariate Gaussian distribution, which is novel in capturing the essence of communities as distributions rather than as discrete entities. This model not only identifies communities within a graph but also optimizes node embeddings by incorporating community-aware high-order proximity, which facilitates a deeper understanding of the graph's structure. Experimental results on multiple real-world datasets demonstrate ComE's superiority in improving graph visualization and outperforming state-of-the-art methods in community detection and node classification tasks. The approach's efficiency is underscored by its linear complexity with respect to the size of the graph, making it a scalable solution for large datasets.

% (Community Detection Based on DeepWalk Model in Large-Scale Networks)
\textbf{The Community Detection Based on DeepWalk Model in Large-Scale Networks} \citep{chen2020community} proposes a method that combines the network embedding representation method with DeepWalk and a GMM, enhanced by variational inference for community detection in large-scale networks. This approach aims to accurately and efficiently detect community structures by embedding high-dimensional network data into a low-dimensional space, preserving topology information. The DeepWalk model is utilized to learn a dense and continuous representation of vertices, facilitating the reduction of noise and the preservation of intrinsic structural information. The embedded data are then clustered using a variational Bayesian GMM, which can automatically determine the number of communities through variational inference, overcoming the limitations of traditional clustering methods that require pre-defined community numbers. Experiments on the DBLP dataset demonstrate that this method not only effectively discovers communities within large-scale networks but also reveals the organizational characteristics within these communities. The model's significant advantage lies in its ability to automatically adjust the number of clusters to match the actual community structure, offering a scalable and insightful approach to understanding complex network dynamics.

\subsubsection{Hidden Markov Models (HMM)}

\textbf{Hidden Markov Models} \citep{Rabiner1989}serve as a prominent probabilistic framework for analyzing sequential data, finding applications in diverse fields like speech recognition and natural language processing. The fundamental assumption of HMM revolves around the concept of a Markov process, wherein the system possesses hidden states, and the observed data is contingent upon the present hidden state. The model comprises essential components, including transition probabilities delineating the shifts between states, emission probabilities governing the generation of observations, and initial state probabilities. The intricacies of HMM necessitate specialized inference methods for their effective utilization. Two commonly employed methods are the Viterbi method, introduced by Forney in 1973, and the Forward-Backward method, proposed by Devijver in 1985. These methods play a pivotal role in extracting meaningful information from sequential data by deciphering the underlying hidden states and decoding the intricate relationships between them. Thus, HMM, with their robust mathematical foundations and efficient inference mechanisms, stand as a cornerstone in the analysis of sequential data across various domains \citep{Devijver1985}.

%(Community detection model for dynamic networks based on hidden Markov model and evolutionary algorithm )
\textbf{HMM-MODCD model} \citep{abbood2023community} applies a multi-objective evolutionary method integrated with a hidden Markov model to detect evolving communities in dynamic networks. This approach decomposes the dynamic community detection challenge into three essential components: intra-community connections, inter-community connections, and community evolution. The HMM-MODCD model uses the Viterbi method for temporal analysis, providing a framework that captures the evolution of community structures over time. Experimental evaluations on synthetic and real-world dynamic networks demonstrate the model's superiority over existing methods, highlighting its ability to accurately trace community evolution while maintaining high-quality community detection across different network states. The HMM-MODCD model's novel integration of evolutionary methods with hidden Markov models presents a significant advancement in dynamic community detection, offering a scalable and effective solution for analyzing complex, evolving networks.

\subsubsection{Bayesian-based Method}

\textbf{Bayesian networks} \citep{Pearl1988}, distinguished as graphical models, provide a structured framework for depicting probabilistic relationships among variables through directed acyclic graphs. In these graphs, nodes symbolize variables, while edges delineate the conditional dependencies existing between these variables. The intrinsic strength of Bayesian networks lies in their ability to capture intricate relationships and dependencies, making them valuable for diverse applications ranging from artificial intelligence to epidemiology.
Within the graphical representation, each node is linked with a conditional probability distribution that precisely characterizes the likelihood of a node given its parent nodes. This probabilistic framework empowers Bayesian networks to model complex systems where variables interact in a dynamic and interdependent manner. The elegance of Bayesian networks lies not only in their capacity to model relationships but also in their utility for inference.
Performing inference in Bayesian networks involves extracting meaningful conclusions or predictions from the model. Several techniques are employed for this purpose, including variable elimination, belief propagation, and Monte Carlo methods like Markov Chain Monte Carlo (MCMC). These methodologies allow researchers and practitioners to navigate the complexity of the network, unraveling hidden patterns and making informed decisions based on probabilistic reasoning.
The versatility of Bayesian networks, coupled with their intuitive graphical representation and powerful inference methods, positions them as indispensable tools for modeling uncertainty and making informed decisions in fields as diverse as finance, healthcare, and environmental science. As technology advances, Bayesian networks continue to evolve, adapting to new challenges and expanding their applicability in an ever-changing landscape of data analysis and decision-making \citep{Gallagher2009}.

\textbf{BCD model}
%(Bayesian Community Detection(2012)) 
 \citep{morup2012bayesian} is a nonparametric Bayesian approach designed to discover community structures within networks. Unlike previous models, BCD adheres to an intuitive definition of communities, where vertices within clusters have many internal links and comparatively fewer links to vertices in different clusters. The model employs a MCMC procedure for inferring the community structure and introduces a novel parameter to gauge the extent of community cohesion, which is learned from data. This parameter helps in distinguishing between internal and external link densities of communities, significantly enhancing the model's ability to predict missing links in both synthetic and real-world networks. The BCD model's efficacy in detecting communities that closely match the actual distribution of links surpasses existing approaches, highlighting the importance of explicitly modeling community structures in network analysis. A Matlab toolbox for the BCD inference procedure has been made available, facilitating the model's application to various networks. This approach not only outperforms existing models in terms of link prediction accuracy but also provides deeper insights into the underlying community structure of networks.

\textbf{NEGCD model}
%(A network embedding-enhanced Bayesian model for generalized community detection in complex networks)
 \citep{he2021network} integrates network embedding techniques with a Bayesian probabilistic model for detecting generalized communities in complex networks. By utilizing the adjacency matrix alongside network embedding—dense, continuous vector representations of nodes in low-dimensional space—this model aims to overcome the limitations of traditional methods in handling noise and redundant information in network topology. The NEGCD model operates by inferring community memberships through a Bayesian treatment of model parameters and employs an efficient variational inference method for community detection. Demonstrating superior performance on both synthetic and real-world networks, the NEGCD model not only accurately detects community structures but also provides insights into the underlying network mechanisms, effectively addressing both assortative and disassortative structures. This approach stands out for its ability to describe generalized communities meaningfully, highlighting its potential in advancing network analysis applications.

\textbf{BCDC}
%(Bayesian community detection for networks with covariates)
 \citep{shen2022bayesian} introduces a novel Bayesian approach to detect communities in networks by incorporating additional node or edge covariate information into the SBM. The primary technique utilized is a covariate-dependent random partition prior, which significantly incorporates the covariate information into the prior distribution of cluster memberships, enhancing the model's flexibility in estimating uncertainties of all parameters, including community memberships. This model stands out for its capability to infer the number of communities directly from the data, without prior specification. It shows improved performance in community detection tasks across both dense and sparse networks with varying covariate types (categorical or continuous), as demonstrated through efficient MCMC method simulations. The BCD-Cov model's ability to integrate covariate data into the community detection process not only provides a more nuanced understanding of the network's structure but also offers an innovative approach to the analysis of complex networks, marking a significant advancement in the field of network analysis.

\subsubsection{Stochastic Block Model-Based Methods}

\textbf{SBM}
%(Stochastic block-models: First steps)
 \citep{mossel2012stochastic} represent a groundbreaking method for the analysis of social networks. By partitioning network actors into distinct groups or blocks, SBM introduces a stochastic extension to traditional deterministic blockmodels, accommodating the inherent variability in relational data. This innovative model illuminates the structural intricacies of social networks by categorizing nodes into comprehensive subsets based on mutual exclusivity and exhaustiveness, where the connections between nodes are influenced by their respective block affiliations. SBM is adept at unraveling the complex web of relationships within networks, offering a sophisticated tool for the methodological analysis of relational data—a domain previously hampered by the scarcity of advanced analytical techniques. The model's versatility extends to modeling reciprocal ties, thereby broadening its descriptive and inferential utility. Through illustrative examples and specialized estimation methods, SBM provide a potent statistical methodology, poised for widespread application in the exploration of relational data. This methodological leap forward facilitates a nuanced understanding of network structures, bridging the gap between global structural descriptions and data variability. 

\textbf{LSBM}
%（Community Detection in the Labelled Stochastic Block Model(2012)）
 \citep{heimlicher2012community} for community detection, explored by Simon Heimlicher, Marc Lelarge, and Laurent Massoulié in 2012, extends the traditional stochastic block model by incorporating multiple types of interactions (or labels) between nodes, offering a refined analysis of network structures. This model partitions nodes into communities and assigns labels to interactions, enabling the differentiation of connections not just by their existence but by their nature or type. Focusing on a two-community scenario, the authors introduce a threshold condition for the feasibility of community reconstruction that correlates with the true partitioning of the network. They provide theoretical and numerical evidence supporting this threshold, highlighting a phase transition in belief propagation's effectiveness for community detection based on label information. This study advances our understanding of network dynamics by integrating interaction types into community detection, promising enhanced accuracy in identifying and interpreting complex community structures within networks. The LSBM represents a significant methodological innovation, enabling the analysis of networks where the nature of interactions plays a crucial role in defining community boundaries.

\textbf{Spectral Partition algorithm}
%(Accurate Community Detection in the Stochastic Block Model via Spectral Algorithms(2014)
 \citep{yun2014accurate} addresses community detection in the SBM with a finite number of communities, where the network consists of a random graph with nodes connected within and across communities at different probabilities. This model, particularly focusing on sparse networks where the connection probability p within communities grows faster than the logarithm of the network size n, leverages spectral methods to classify nodes into their respective communities. The authors demonstrate that, under certain conditions related to the network's connectivity and the sizes of its smallest communities, the spectral methods guarantee with high probability that the number of misclassified vertices remains below a small fraction s of the total network size as n grows large. This result not only corroborates the feasibility of spectral methods in accurately detecting community structures in SBM but also suggests the optimality of such methods under more general conditions than previously established. The study contributes significantly to understanding the capabilities and limits of spectral methods in community detection, proposing a practical and computationally efficient approach to unraveling complex network structures.

\textbf{Multi-stage Maximum Likelihood algorithm}
%(A scalable community detection algorithm for large graphs using stochastic block models)
 \citep{peng2015scalable} offers a scalable solution for community detection in large graphs using SBM. Aimed at addressing the limitations of traditional inference methods in handling large-scale data, this approach optimizes for speed, scalability, and quality of results. The method employs a coordinate descent method with approximations to streamline computations, ensuring that its runtime is linear with respect to the number of edges. A notable feature is its parallelization through a message-passing framework, enabling significant speedup without compromising accuracy as the processor count increases. By initializing nodes into many tiny communities and employing a multi-stage strategy for gradually constructing larger communities, the method ensures high-quality detection results. Tested on both benchmark and real-world graphs, it outperforms traditional community detection methods, demonstrating its effectiveness in uncovering meaningful community structures within networks.

\textbf{RSBM}
%(A Regularized Stochastic Block Model for the robust community detection in complex networks) 
 \citep{lu2019SBM} extends the degree-corrected stochastic block model by incorporating regularization on nodes' internal degree ratios. This enhancement is designed to guide the inference methods towards recovering either assortative or disassortative structures within a network, depending on a single parameter's value. Traditional inference methods often struggle to identify weak assortative structures, converging instead to disassortative partitions that, while locally optimal, may not represent the sought-after community structures. By imposing constraints on the internal degree ratios within the objective function, the RSBM effectively directs the inference process, ensuring that assortative or disassortative structures are reliably identified as intended. Experimental results demonstrate the model's efficacy in both synthetic and real-world networks, showing that it not only consistently identifies the correct community structures but does so more rapidly and reliably than when using the degree-corrected model alone. This model's introduction represents a significant advancement in community detection methodologies, offering enhanced control over the type of community structures identified in complex networks.

\textbf{DCD algorithm}
%（A distributed community detection algorithm for large scale networks under stochastic block models）
 \citep{wu2023distributed} is a novel approach tailored for large-scale network analysis. It focuses on the spectral clustering of network nodes distributed across a master server and several worker servers. This method distinguishes itself by first conducting spectral clustering on a subset of pilot nodes on the master server to identify pseudo centers, which are then used to facilitate community detection on worker servers without requiring further iterative processes. This results in a method that is not only computationally efficient but also reduces communication costs and storage requirements, as it avoids the need to use the entire adjacency matrix. A Python package, DCD, supports the method's implementation on a Spark system, demonstrating its practical applicability through experiments on synthetic and real-world datasets. The DCD algorithm's advantages include its low communication overhead, the absence of iterative computations on worker servers, and its scalability, making it an effective tool for detecting community structures in large networks.

\subsubsection{LDA}

% \citep{Blei2003}:}

\textbf{LDA} \citep{Blei2003} stands as a powerful generative probabilistic model extensively employed in natural language processing for the purpose of topic modeling. This innovative approach posits that documents arise from a composite of latent topics, with each topic being defined by a distinctive probability distribution over the words contained within. The primary objective of LDA is to uncover and comprehend these latent topics based on the observable content of documents. The process of inference in LDA is commonly executed through two main techniques: variational inference and Gibbs sampling. Variational inference involves approximating the complex posterior distribution of latent variables, optimizing a lower bound on the model's likelihood. On the other hand, Gibbs sampling is a Markov Chain Monte Carlo method that iteratively samples latent variables conditional on the observed data, converging towards a representative distribution. Both methodologies contribute significantly to the effectiveness of LDA, enabling it to unveil hidden thematic structures within textual data, thus enhancing our understanding of the underlying patterns in language.

\textbf{The SSN-LDA model}
%(An LDA-based community structure discovery approach for large-scale social networks)
 \citep{zhang2007lda} employs an LDA-based hierarchical Bayesian algorithm for community discovery in large-scale social networks. Unlike traditional methods, SSN-LDA models communities as latent variables, defined as distributions over the social actor space, with the unique advantage of requiring only topological information as input. The model's strength lies in its ability to uncover community structures by processing network topologies, demonstrating promising results on collaborative networks like CiteSeer and NanoSCI. SSN-LDA is distinguished for its novel approach of treating communities as mixtures over latent spaces, which simplifies the complexity of community detection and enhances the interpretability of the results. The model is evaluated based on its effectiveness in revealing coherent community structures within networks, showcasing its potential for broader applications beyond the tested datasets. Its innovation in leveraging latent Dirichlet Allocation for community discovery marks a significant advance in the field, offering a scalable and insightful methodology for analyzing complex social networks.

\textbf{Bayesian probabilistic model}%(A model-based approach to attributed graph clustering)
 \citep{xu2012model} designed for the efficient clustering of attributed graphs. This approach stands out by seamlessly integrating structural and attribute information of nodes without relying on artificial distance measures. Central to the model is its capability to simultaneously account for the structural connections and attribute similarities among graph nodes, thereby facilitating a more natural and principled clustering process. The transformation of the clustering task into a probabilistic inference problem, addressed with an effective variational method, showcases its superiority over traditional distance-based methods in achieving higher clustering quality across various datasets. By adopting a model-based perspective for attributed graph clustering, the method adeptly balances structural and attribute data, leading to more coherent community detection in complex networks. This innovative strategy underscores the advantages of model-based techniques in enhancing graph analysis, particularly in their capacity to cohesively process multifaceted information within a unified probabilistic framework.

\textbf{The LDA-LPA model}
%(A Community Detection Method Based on Topic Model)
 \citep{Wang2021LDA} presents a novel community detection approach grounded in topic modeling, specifically tailored for complex networks lacking attribute information. Unlike conventional topic model-based community detection methods, which typically depend on node or edge attributes like text or images, LDA-LPA solely requires structural network data, thus broadening its applicability. This model conceptualizes communities as distinct topics and represents network nodes as documents through a random walk process. Applying the LDA model to this corpus allows for the inference of each node's community affiliation based on topic distribution. Initial community assignments are refined using a label propagation algorithm, further enhancing the precision of community detection. The model demonstrates superior performance over several benchmark methods across both real and synthetic datasets, particularly achieving perfect community detection scores in certain cases. The LDA-LPA model's introduction marks a significant advancement in community detection, offering a robust, attribute-independent methodology that promises wide-ranging applicability and efficiency in analyzing complex networks.

\subsubsection{MRF(Markov Random Field) based}

\textbf{NetMRF model}
%(Network-Specific Markov Random Field Approach to Community Detection) 
 \citep{he2018network} pioneers the application of Markov Random Fields for community detection in networks. This model stands out for its unique approach to capturing the structural properties of irregular networks by focusing on the pairwise potential among nodes, rewarding internal edges within communities and penalizing edges between different communities. The NetMRF approach leverages a network-specific belief propagation method for model inference, efficiently identifying community structures while accommodating the network's topology. Experimental results on synthetic benchmarks and real-world datasets underscore the NetMRF model's superior performance in community detection over state-of-the-art methods. This advancement demonstrates the potential of probabilistic graphical models in addressing complex network analysis challenges, particularly in identifying coherent communities in large and intricate networks.

\textbf{MRFasGCN model}
%(Graph convolutional networks meet markov random fields: Semi-supervised community detection in attribute networks MRFasGCN) 
 \citep{jin2019graph} innovatively combines Graph Convolutional Networks (GCN) and MRF for semi-supervised community detection in attribute networks. By leveraging the strengths of both GCN and MRF, this model offers a comprehensive approach that utilizes network topology and node semantic information within an end-to-end deep learning architecture. The extensive experiments across various large benchmark problems demonstrate MRFasGCN's superiority over state-of-the-art methods in terms of performance and scalability. The model's unique integration facilitates a nuanced analysis of complex networks, marking a significant advancement in community detection methodologies by enabling the exploration of both structural and attribute data to uncover community structures efficiently and accurately.

\textbf{The ModMRF method}
%(ModMRF: A modularity-based Markov Random Field method for community detection)
 \citep{jin2020mod} introduces a novel modularity-based Markov Random Field approach for community detection in complex networks. By formalizing modularity as an energy function within a Markov Random Field framework, ModMRF innovatively addresses the challenge of detecting community structures in networks. It employs belief propagation for inference, achieving a principled way to ascertain the number of communities, a task at which traditional modularity optimization methods falter by often partitioning networks into an excessively large number of communities. ModMRF's distinctiveness lies in its statistical model formulation of modularity, enabling the application of various statistical inference methods for optimization. This method outperforms existing methods in terms of accuracy and efficiency on both real-world and synthetic datasets, demonstrating its capability to discover meaningful community partitions while effectively managing the computational complexity associated with large networks.

\subsubsection{Non-negative Matrix Factorization(NMF) based method}

\textbf{Non-negative Matrix Factorization}
%(Algorithms for non-negative matrix factorization) 
 \citep{lee2000algorithms} serves as a valuable tool for the decomposition of multivariate data. Unlike traditional factorization methods, NMF constrains the components to be non-negative, enabling a parts-based representation of data. This model stands out by providing two distinct multiplicative methods for its implementation, differing in their optimization criteria: one minimizes the Euclidean distance, while the other targets the Kullback-Leibler divergence. Through rigorous analysis, both methods are shown to guarantee monotonic convergence, akin to the Expectation-Maximization method, offering a novel perspective on data approximation. The intrinsic value of NMF lies in its ability to reveal hidden structures within datasets, making it especially applicable in fields like image processing, text mining, and bioinformatics, where interpretability and the extraction of meaningful components are crucial. This work not only introduces a significant advancement in unsupervised learning but also broadens the horizon for future research in data analysis and representation.

\textbf{NMF-based method}
%(Clustering complex networks and biological networks by non-negative matrix factorization with various similarity measures) 
 \citep{wang2008clustering} for detecting community structure in complex networks introduces a versatile approach to community detection across various networks, including biological systems. This method employs several similarity measures, such as diffusion kernel similarity and shortest path-based similarity, to evaluate connections within a network, and then applies NMF to identify distinct community structures. Particularly, the diffusion kernel similarity measure is highlighted for its effectiveness in large biological networks, showcasing the ability of the NMF method to uncover biologically meaningful functional modules. Comparative analyses against other methods demonstrate the NMF method's superior performance in both synthetic and real-world datasets, emphasizing its capacity to discern intricate community dynamics efficiently. This approach not only expands the applicability of NMF in the field of network analysis but also underscores the importance of selecting appropriate similarity measures to enhance the detection of community structures within complex networks.

\textbf{SymNMF model}
%(Symmetric nonnegative matrix factorization for graph clustering) 
 \citep{kuang2012symmetric} is an innovative framework designed for graph clustering. This approach extends traditional NMF techniques by utilizing a symmetric matrix that contains pairwise similarity values between data points, thus adapting NMF for applications in graph clustering. Unlike standard NMF, SymNMF operates on a similarity matrix that is not necessarily nonnegative, which allows for a broader application in capturing the inherent structure of complex networks. Through the introduction of SymNMF, the authors have provided a robust method that inherits the interpretability and advantages of NMF while making it applicable to graph data. This method proves particularly effective in identifying community structures within graphs, offering a novel and insightful perspective on data clustering. The development of a Newton-like method for efficient computation further enhances the practical utility of SymNMF, making it a significant contribution to the field of data analysis and network science.

\textbf{Nonnegative Matrix Factorization based on Node Centrality (NCNMF)} \citep{su2023}. The paper addresses the limitations of existing NMF-based methods by incorporating higher-order proximity information and considering the importance of nodes in community detection. The proposed method utilizes a new similarity measure that captures the topological structures between nodes and introduces node centrality and Gini impurity to promote sparsity in community memberships. The NCNMF method combines matrix factorization and graph properties, ensuring scalability and interpretability while effectively detecting communities. Experimental results demonstrate its superior performance compared to state-of-the-art methods.

\textbf{PODCD}
%(PODCD: Probabilistic overlapping dynamic community detection)  
 \citep{bahadori2021podcd} presents a cutting-edge strategy for discerning community affiliations within dynamic networks, uniquely accounting for the temporal evolution of these communities. Distinct from conventional approaches that overlook or independently analyze temporal data, PODCD employs a probabilistic graphical framework to cohesively model the dynamism of community interactions over time. This technique is particularly proficient in identifying intersecting community memberships, allowing for nuanced memberships across diverse communities. At its foundation, PODCD utilizes an enhanced version of Non-Negative Matrix Factorization, incorporating temporal regularization to dynamically adjust memberships based on historical trends. Through rigorous testing on a variety of datasets, PODCD has proven its efficacy, surpassing other methods in accurately mapping community evolution with notable precision and temporal coherence. This method enriches our comprehension of community dynamics, offering a scalable solution for the examination of evolving networks.

\textbf{The Non-Negative Matrix Factorization for Detecting Communities in Networks method}
%(An Algorithm Based on Non-Negative Matrix Factorization for Detecting Communities in Networks)
 \citep{huang2024algorithm} introduces a innovative  approach to community detection within sparse network communities by utilizing NMF. This method, aimed at overcoming the limitations faced by conventional NMF-based methods in sparse networks, leverages local feature vectors to better represent the original network's topological features through regularization matrices. The essence of this approach lies in its capacity to handle the sparsity and noise typical of real-world network data, enhancing the feature expression capabilities and enabling a more accurate detection of community structures. The authors demonstrate the superiority of their method over standard NMF-based techniques through extensive experimentation on both simulated and real-world networks. The method stands out by resolving the issue of localized feature vectors, thus achieving higher accuracy in identifying community structures. However, like any method, it has its limitations, particularly in the need for further reduction of algorithmic complexity and adaptive techniques for estimating community numbers. This method marks a significant advancement in the field of network structure analysis, offering a robust tool for unveiling the intricate community dynamics inherent in complex networks.

%%%%%%%%%%5
%%% 小赖做到了这里，到达了MRF(Markov Random Field) based:；后面大概还有1/3

%%%% 继续做

\subsection{Deep Learning Approaches}

Deep learning models have shown promising results in various domains, including community detection in networks. These approaches leverage neural networks to capture complex patterns in network structures and node attributes. Here are some deep learning models and methods commonly used for community detection

\subsubsection{GCN}

\textbf{GCN} \citep{Kipf2016GCN} represent an innovative extension of traditional convolutional neural networks, specifically designed to process data that is structured as graphs. By ingeniously leveraging the inherent graph structure, GCN is adept at aggregating and propagating information across the neighborhoods of nodes. This unique capability allows them to excel at identifying communities within a network, effectively detecting and grouping nodes based on the intricate patterns of connectivity that define their interactions. The application of GCN-based models transcends the boundaries of simple networks, showcasing remarkable effectiveness in unraveling community structures within a diverse array of networks. Whether it's the complex web of relationships in social networks, the intricate citation connections among academic papers, or the sophisticated interaction networks within biological systems, GCN has proven to be an indispensable tool in the analysis of networked data. Their ability to discern the subtle nuances of community affiliation within these varied networks underscores the transformative potential of GCN in advancing our understanding of complex systems.

%(Attributed graph clustering via adaptive graph convolution)
\textbf{Attributed Graph Clustering via Adaptive Graph Convolution method} \citep{zhang2019attributed} heralds a new era in clustering methodologies by integrating node attributes and graph topology through the innovative use of adaptive graph convolution. This method stands out by employing graph convolution of varying orders to aptly capture the overarching structure of communities, subsequently refining the clustering process based on the graph's unique characteristics. This dual-focus on high-order graph convolution and the strategic selection of convolution depth ensures that node features are smoothly integrated, thus significantly enhancing the quality of clustering outcomes. Through empirical analysis across a spectrum of datasets, this approach demonstrates unparalleled performance, significantly outpacing both traditional and contemporary clustering models in accuracy, comprehensiveness, and efficacy. The methodology's core strength lies in its adaptive nature, efficiently navigating the complexities of real-world network data to deliver insights with profound implications for community detection and network analysis.

\textbf{Unsupervised Learning for Community Detection in Attributed Networks Based on GCN}
%(Unsupervised Learning for Community Detection in Attributed Networks Based on Graph Convolutional Network)
 \citep{wang2021unsupervised} presents an innovative unsupervised learning framework that utilizes GCN for community detection in attributed networks without requiring prior label information. The core innovation integrates a label sampling model with GCN to uncover community structures effectively by leveraging both topological and attribute information. The label sampling model constructs a balanced training set by identifying structural centers and expanding neighboring nodes for GCN model training. This approach not only boosts community detection accuracy but also pioneers the use of GCN in unsupervised network analysis. Through extensive experiments on various real-world networks, the method demonstrates superior community detection performance over state-of-the-art techniques, highlighting the potential of combining unsupervised learning with GCN for advanced network analysis tasks.

\textbf{CLARE:Community Locator and Rewriter}
%The CLARE algorithm (CLARE: A semi-supervised community detection algorithm)
 \citep{wu2022clare}introduces a novel semi-supervised framework for community detection in networks. The method innovatively combines the processes of locating potential communities through subgraph matching and refining them using deep reinforcement learning. Specifically, it employs a Community Locator to identify potential communities by matching subgraphs similar to provided training examples, and a Community Rewriter to adjust these communities, optimizing their structure based on global patterns. This dual-component strategy allows CLARE to accurately identify and refine community structures within networks, leveraging both local subgraph similarities and global structural insights. Experimental results demonstrate CLARE's effectiveness in achieving superior performance in community detection tasks across multiple real-world datasets, showcasing its ability to significantly outperform existing state-of-the-art methods in terms of accuracy and robustness, particularly in complex, noisy, or large-scale network environments.

\textbf{CPGC:Community detection based on community perspective and graph convolutional network}
%The CPGC method(Community detection based on community perspective and graph convolutional network)
 \citep{liu2023community} integrates the Bernoulli–Poisson model with GCN enhancements for community detection in attributed networks. This unsupervised method addresses the challenge of detecting overlapping communities and the limitation of existing methods that separate representation learning from clustering. CPGC combines representation learning and clustering, employs a modified graph convolution to improve node representation distinguishability, and introduces a novel community perspective similarity to utilize community-level information effectively. The method demonstrates state-of-the-art performance in identifying both nonoverlapping and overlapping communities across various real-world networks, highlighting its ability to leverage attribute and structural information without requiring prior label information. However, it also encounters challenges with computational complexity and scalability, underscoring the need for optimization in handling large-scale networks.

\subsubsection{Graph Autoencoders}

\textbf{Graph Autoencoder (GAE)} \citep{Grover2019,Kipf2016VAE} represent a pivotal advancement in neural network architectures tailored for the nuanced analysis of graph-structured data. These models embark on a mission to distill the essence of complex network structures by learning compact, low-dimensional representations of individual nodes while meticulously preserving the intricate web of connections that define the graph's topology. Through a process of encoding and decoding, GAE ingeniously maps the network's structural information into a latent space, effectively capturing the essence of community dynamics. Among the pantheon of GAE variants, Variational Graph Autoencoders (VGAE) and Graph Convolutional Autoencoders (GCAE) stand as vanguards, each offering unique strengths in augmenting the performance of community detection tasks. VGAE, with their probabilistic framework, introduces an element of uncertainty into the latent space, enhancing the robustness of community detection by accounting for the inherent variability in network data. On the other hand, GCAE, by integrating the principles of graph convolutional networks, empowers the auto-encoder with the ability to exploit local graph structure, fostering a more nuanced understanding of community dynamics. As such, the advent of GAE and their progeny signifies a paradigm shift in the realm of network analysis, promising to unravel the intricate tapestry of communities lurking within the vast expanse of interconnected data.

\textbf{VGAE:Variational Graph Auto-Encoder}
%The VGAE model(VGAE Variational graph auto-encoders)
 \citep{Kipf2016VAE} is an unsupervised learning framework for graph-structured data that extends the variational auto-encoder (VAE) concept. Utilizing latent variables, VGAE leverages a GCN encoder and a simple inner product decoder to learn interpretable latent representations for undirected graphs. This model is distinctive for its ability to naturally incorporate node features, which significantly enhances predictive performance on several benchmark datasets, particularly in link prediction tasks for citation networks. Unlike most existing models for unsupervised learning on graph data, VGAE shows competitive results by achieving improved link prediction accuracy through the integration of node attributes into the learning process. However, the reliance on a Gaussian prior with an inner product decoder, which tends to push embeddings away from the zero-center, is identified as a potential limitation. Future work aims to explore more suitable prior distributions and more flexible generative models to further enhance the model's scalability and performance.

\textbf{SEDCN:Structure Enhanced Deep Clustering Network via a Weighted Neighbourhood Auto-Encoder}
%The SEDCN method(Structure enhanced deep clustering network via a weighted neighbourhood auto-encoder)
 \citep{bai2022structure} introduces a novel framework for structural deep clustering that integrates semantic and structural representations for clustering tasks. The core innovation of SEDCN lies in its use of a weighted neighbourhood auto-encoder (wNAE) to enhance the structural data representation, ensuring that the structural information does not vanish during the learning process. This method addresses the over-smoothing problem often encountered with GCN by providing a dual mechanism that combines the advantages of auto-encoders for semantic information learning and GCN for structural representation. The SEDCN method demonstrates superior clustering performance across various datasets, showcasing its ability to effectively capture and utilize both semantic and structural information for improved clustering outcomes. Additionally, the introduction of a joint supervision strategy ensures the coherent optimization of the wNAE and GCN modules alongside the clustering assignment, further enhancing the model's performance and robustness.

\textbf{SSGCAE:Semi-Supervised Graph Convolutional Autoencoder for Overlapping Community Detection}
%The SSGCAE method( Semi-supervised overlapping community detection in attributed graph with graph convolutional autoencoder) 
 \citep{he2022semi} innovatively combines a GCAE with modularity maximization and semi-supervision modules to tackle the challenge of detecting overlapping communities in attributed graphs. This semi-supervised approach effectively fuses link and attribute information while integrating prior knowledge to enhance detection accuracy. The key technique, GCAE, serves as the backbone, seamlessly integrating structural and attribute data through a unified framework that optimizes community membership representation. SSGCAE demonstrates superior performance over existing methods in extensive experiments on both synthetic and real-world datasets, establishing its effectiveness in identifying meaningful community structures. This advancement is particularly noteworthy for its ability to handle overlapping communities, a common scenario in complex networks. The approach's innovation lies in its semi-supervised learning model, leveraging prior information for improved detection outcomes, and addressing the challenges of community overlap and dynamic network structures.

\textbf{CSEA:Core Structure Extraction Algorithm using Variational Autoencoder for Community Detection}
%The CSEA method(A novel network core structure extraction algorithm utilized variational autoencoder for community detection)
 \citep{fei2023novel} pioneers in integrating the K-truss method with a variational auto-encoder to enhance community detection in networks. By focusing on the network's core structure, CSEA leverages the K-truss method to identify densely connected node groups, generating a similarity matrix that captures core structural features. This matrix is then refined through a variational auto-encoder, reducing dimensionality while preserving essential information. The method distinguishes itself by combining structural insights with advanced unsupervised learning, aiming for accurate community segmentation. CSEA's effectiveness is validated across diverse real-world networks, showing notable improvements over traditional methods, particularly in dense, complex networks. However, challenges remain in optimizing algorithmic complexity and automatically determining community numbers, highlighting areas for future advancement in network analysis.

\textbf{AWBA:Attentional-Walk-Based Autoencoder for Community Detection}
%The AWBA algorithm(An attentional-walk-based autoencoder for community detection) 
 \citep{guo2023attentional} leverages an encoder-decoder structure incorporating attention mechanisms and a novel random walk strategy to enhance community detection in complex networks. This method emphasizes capturing both low-order and high-order relationships among nodes by integrating attention coefficients and community membership into the transition probabilities of random walks. The significant contribution of AWBA is its ability to discover communities with vague boundaries by effectively utilizing the nonlinear relationships between nodes. Experimental results on both synthetic and real-world datasets demonstrate AWBA's superior performance in accurately identifying communities, outperforming baseline methods. The approach is innovative in combining attention layers within the encoder for learning node influences and a community-aware random walk in the decoder, addressing the limitations of traditional auto-encoders that focus primarily on adjacency matrix reconstruction.

% \subsubsection{Graph Generative Models}
% % \citep{Bojchevski2018,You2018}:}

% \Generative models, particularly those tailored for graph data like Graph Variational Autoencoders (GraphVAE) and Graph Recurrent Neural Networks (GraphRNN), are pivotal in capturing the inherent distribution of network data, thereby facilitating the creation of novel instances closely resembling the original network. These advanced frameworks excel in generating authentic graph structures, which serve as a foundation for various analytical tasks. One such application is community detection, wherein the generated structures are meticulously analyzed to unveil inherent communities within them. This process not only sheds light on the organizational patterns embedded in the data but also offers insights into the intricate relationships between different entities within the network. By leveraging graph generative models, researchers and analysts can gain a nuanced understanding of complex network dynamics, transcending traditional methods and opening avenues for deeper exploration and discovery. Thus, the versatility of graph generative models lies in their ability to uncover meaningful structures, unravel hidden patterns, and enhance our comprehension of the underlying dynamics governing interconnected systems.

\subsubsection{Graph Attention Networks (GAT)} 
% \citep{Velickovic2017}:}

\textbf{GAT} \citep{Velickovic2017} revolutionize the landscape of graph-based modeling by harnessing attention mechanisms, which dynamically allocate weights to neighboring nodes. This pivotal innovation empowers GAT models to discern and prioritize pertinent information during the intricate process of message passing. Through this nuanced mechanism, GAT effectively encapsulates the essence of community structures inherent in complex networks. By assigning heightened attention to nodes within the same community while dampening the significance of nodes from disparate communities, GAT encapsulates the intricate web of interconnections that define community boundaries. The efficacy of this approach shines through in various community detection tasks, where GAT consistently demonstrates remarkable performance and robustness. Its ability to adaptively navigate through the network, focusing on salient regions while disregarding noise, underscores its utility as a versatile tool for analyzing and understanding the intricate fabric of network dynamics. In essence, GAT represents a pivotal advancement in graph-based modeling, offering insights and solutions to the complex challenges of community detection and network analysis.

\textbf{HiAN:Hierarchical Attention Network for Attributed Community Detection of Joint Representation}
%The HiAN method(Hierarchical attention network for attributed community detection of joint representation)
 \citep{zhao2022hierarchical} introduces a novel framework for detecting communities within attributed networks. By employing a unique hierarchical attention mechanism, HiAN excels in integrating both the structural and attribute data of graphs to model the interactive information and interpretability behind attributed graphs efficiently. This approach leads to the simultaneous optimization of node representation and community detection, facilitated by a self-training clustering component that enhances detection accuracy through mutual benefit between representation learning and community assignment. Experimental validation on various real-world datasets demonstrates HiAN's superior performance against several state-of-the-art baselines, highlighting its ability to effectively and interpretably identify community structures in attributed networks. Despite its advantages, the method's computational complexity and the challenge of automatic community number estimation remain areas for future improvement.

\textbf{GEAM:Graph-Enhanced Attention Model}
%GEAM(A graph-enhanced attention model for community detection in multiplex networks)
 \citep{wang2023graph} for community detection in multiplex networks is designed to encode intra-layer topologies and merge inter-layer semantics efficiently. By utilizing global information within each network layer, GEAM introduces a novel layer contrastive learning module for encoding local and global information, a self-attention adaptive fusion mechanism for fusing multiple layers' encodings, and an edge density-driven community detection module for identifying communities with strong modular structures. Experimental results on both synthetic and real-world datasets demonstrate GEAM's superiority over existing methods, showcasing its ability to provide more accurate community detection in multiplex networks. The method's effectiveness is attributed to its comprehensive approach to learning node representations and optimizing community structures, highlighting its potential for advanced network analysis tasks.

\textbf{The Overlapping Community Detection Using Graph Attention Networks method}
%(Overlapping community detection using graph attention networks) 
 \citep{sismanis2023overlapping} innovates in the domain of community detection in graphs by incorporating GAT. This approach enhances the identification of overlapping communities by employing attention mechanisms to focus on significant nodes within a graph's neighborhood, distinguishing the method from traditional approaches that treat all nodes equally. The integration of GAT allows the model to adaptively assign importance to nodes based on their structural position and connections, facilitating a more nuanced understanding of community structures. Experimental results demonstrate the method's effectiveness in accurately detecting overlapping communities, highlighting its potential for application in complex network analyses. This method represents a significant advancement in community detection, providing a more detailed and flexible approach to understanding the multifaceted relationships within networks. However, the complexity of the attention mechanisms and the need for extensive computational resources for large-scale networks indicate areas for further optimization and research.

\subsubsection{Graph representation learning}   % \citep{Hamilton2017,Xu2018GNN}:}

\textbf{CDFG:Community Detection Fusing Graph Attention Network}
%The CDFG method(A unified framework for community detection and network representation learning)
 \citep{tu2018unified} innovatively marries the principles of auto-encoders and graph attention networks to enhance the detection of communities within attributed graphs. By employing both attribute feature learning through auto-encoders and structural feature learning via graph attention networks, CDFG adeptly incorporates high-order neighborhood information. This method introduces a feature fusion module for initial and secondary fusion of features, along with the incorporation of self-supervision to refine community detection. Demonstrated across various datasets, CDFG's performance superiority in community detection emphasizes its capacity to effectively balance the significance of structural closeness and attribute similarity within communities. Nevertheless, challenges in computational efficiency and determining the number of communities automatically are areas identified for future improvement.

\textbf{The vGraph method}
%(A generative model for joint community detection and node representation learning)
 \citep{sun2019vgraph} introduces a probabilistic generative model for simultaneously conducting community detection and node representation learning in graphs. The innovation of vGraph lies in its ability to model each node as a mixture of communities and each community as a multinomial distribution over nodes, thus allowing for collaborative learning of community membership and node representations. By leveraging a designed variational inference method and incorporating a smoothness regularization term, vGraph ensures that community memberships of neighboring nodes are similar in the latent space, significantly enhancing both tasks' performance. Experimental results on multiple real-world graphs highlight vGraph's efficacy, outperforming many competitive baselines in both community detection and node representation learning. The method's flexibility further extends to hierarchical community detection, demonstrating vGraph's comprehensive approach to analyzing complex network structures.

\textbf{GraphSage} \citep{xu2020ind}, a graph representation learning model that operates on static or non-temporal graphs. GraphSAGE aims to learn lower-dimensional vector embeddings for nodes in a graph by aggregating information from their local neighborhoods. It utilizes a neighborhood sampling technique to efficiently generate training examples and employs a flexible aggregator function to combine information from neighboring nodes. By iteratively updating the node embeddings using an end-to-end trainable framework, GraphSAGE learns to capture both structural and attribute information of nodes in the graph. The experimental results demonstrate that GraphSAGE outperforms existing graph representation learning methods on various downstream tasks, such as node classification and link prediction. Overall, GraphSAGE offers a scalable and effective approach for learning representations on graphs.

\textbf{SCD:Embedding-based Silhouette Community Detection}
%The SCD method(Embedding-based Silhouette community detection) 
 \citep{vskrlj2020embedding} introduces an innovative approach to detecting communities within networks by focusing on clustering network node embeddings—real-valued representations of nodes based on their neighborhoods. This unsupervised technique does not rely on modularity optimization, a common basis for many existing community detection methods, but instead evaluates community structures through clustering node embeddings and assessing their quality using the Silhouette score. The method demonstrated competitive or superior performance compared to established methods like InfoMap and Louvain on both synthetic and real-world networks. Notably, SCD's adaptability to various network learning and exploration pipelines and its application in semantic subgroup discovery for generating human-understandable explanations of communities in a protein interaction network underscore its versatility. Despite its advantages, including applicability to large networks and the ability to work with different node representation learners, SCD faces challenges in computational complexity and determining the optimal number of communities.

\textbf{J-ENC:Joint Embedding for Node representation and Community detection}
%The J-ENC model(Unsupervised learning of joint embeddings for node representation and community detection)
 \citep{khan2021unsupervised} proposes an efficient generative framework for learning node representations and detecting communities within graphs. Utilizing a novel approach, J-ENC leverages community-aware node embeddings, where the learning process ensures that connected nodes are not only spatially closer but also share similar community assignments. This dual objective is achieved through a unique modeling that combines the benefits of variational auto-encoders and graph neural networks. The methodology stands out by learning a single embedding that serves both as the node and its community representation, simplifying the learning process and reducing computational demands. Experiments across various datasets demonstrate J-ENC's ability to outperform existing methods in community detection and node classification tasks, showcasing its versatility and efficiency.

\textbf{DEDW:Dynamic Community Detection Based on Evolutionary DeepWalk}
%The DEDW method(Dynamic Community Detection Based on Evolutionary DeepWalk)
 \citep{qu2022dynamic} introduces an innovative approach for identifying community structures within dynamic networks. By combining the principles of evolutionary clustering with the graph embedding technique DeepWalk, DEDW enhances the representation of dynamic network data, addressing challenges of data sparseness and structural evolution. This method notably advances dynamic community detection by generating node embedding feature vectors that incorporate historical network structure information, thus capturing the temporal smoothness and local changes within communities over time. Experimentation on both synthetic and real-world datasets demonstrates DEDW's ability to accurately mine the evolving characteristics of dynamic communities, significantly improving the accuracy and stability of detection outcomes. However, the method's reliance on parameter tuning and computational efficiency in processing large-scale networks suggests areas for further refinement.

\subsection{Our Methodology – RMS}

\begin{figure}[htbp]
      \centering
      \includegraphics{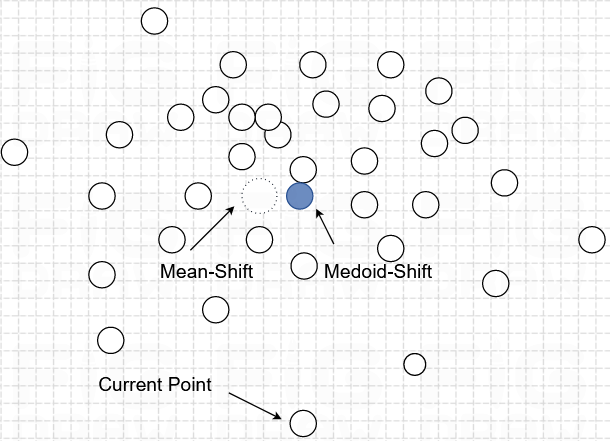}
      \caption{For the current point in the shifting, Medoid-Shift chooses the data point to shift, while Mean-Shift chooses the location to shift}
      \label{fig:1}
  \end{figure}

Derived from the foundational concept of Mean-Shift, Medoid-Shift \citep{Sheikh2007} presents itself as a close sibling, sharing notable parallels in both theory and procedural methodology. Fundamentally, both methods operate by iteratively discerning shifts towards areas of heightened data density, thereby pinpointing cluster centers and autonomously determining the optimal cluster count. However, the pivotal departure, vividly depicted in Figure \ref{fig:1}, lies within the mechanics of these shifts. Whereas Mean-Shift navigates towards a location predicated on the Mean-Shift vector, Medoid-Shift gravitates towards a specific point within its local neighborhood. This nuanced distinction renders Medoid-Shift uniquely suited for applications such as distance-based community detection problems, a capability that eludes Mean-Shift. Consequently, Medoid-Shift emerges as a versatile tool capable of addressing diverse clustering challenges with precision and efficacy \citep{Li2023IC}.

\textbf{Brief Introduction of Medoid-Shift Core Algorithm. } Given an N × N symmetric matrix D(i, j) which is the distance between i and j starting from the point i, an index of point j is calculated as follows:

\begin{equation}
S(i,j) = \sum_{k=1}^{N} D(j,k)\phi D(i,k)
\end{equation}

In the shifting process, the next point to be shifted for point i is determined by selecting point j with the minimum value in $S(i, j)$, where $1 <= j <= n$. By iteratively computing the next point for all current points, a tree traversal technique can be employed to identify unique roots for each point. The number of clusters corresponds to the number of unique roots, and the label of each point can be directly derived from its corresponding unique root.

The RMS algorithm, proposed in this study, exhibits several differences compared to the Medoid-Shift algorithm. Firstly, in Medoid-Shift, the neighborhood is defined by a distance parameter, whereas in RMS, it is defined using the KNN approach. Secondly, when selecting the next point, Medoid-Shift calculates distances between all the points in the current medoid's neighborhood, whereas RMS computes similarities between each point p in the neighborhood of the current medoid and p's KNN.

In the remaining part of subsequent section, we introduce the proposed RMS data preprocessing method. Subsequently, we present the core algorithm of RMS and analyze its advantages over the Medoid-Shift algorithm.

\subsubsection{The Definition of Graph in Social Network. }

Considering a social network denoted as N, it can be represented as a graph $G(V, E)$, where $u \in V$ represents a node, $e \in V $represents another node, and $(u, e) \in E$ represents an edge that signifies a certain type of relationship or intimacy between u and e. The number of edges and nodes in the network are denoted by $|E|$ and $|V|$, respectively. It is important to note that in the context of RMS, only the existence of a connection between the two endpoints is considered, without considering the direction. Therefore, directed graphs are transformed into undirected graphs by aggregating the weights of the directed edges between every pair of endpoints.

\begin{algorithm}[h]  %其中这里面不能有H不然会报错，不过不影响结果
	\caption{Medoid Clustering}%算法名字
	\LinesNumbered %要求显示行号
	\KwIn{$SimM(1..|V|, 1..|V|), k$}%输入参数
	\KwOut{cluster centers, the center points list of final clustering result}%输出
	$SetA \leftarrow {1,2, ..., |V|}$;\\
        $SetB \leftarrow Ø$;\\
        $nextMedoid \leftarrow []$;\\
        \While{True}
        {
        \For{$each i in SetA$}{
        $SetP \leftarrow {i} \cup NN(i)$;\\
        $temp  \leftarrow The point with the largest DL value in the SetP $ ;\\
         $nextMedoid[i]  \leftarrow temp   $ ;\\
        $SetB.add(temp)$;\\
        }
        
        \If{$SetA == SetB$ }{
        \For{$each i in SetA$}{

        $nextMedoid[i] \leftarrow i$
        }
        break;
        }
        $SetA, SetB \leftarrow SetB, Ø$
        }
        \textbf{Return:} centers, nextMedoid
        \label{algor:1}
\end{algorithm}

\begin{algorithm}[h]  %其中这里面不能有H不然会报错，不过不影响结果
	\caption{Label Assignment}%算法名字
	\LinesNumbered %要求显示行号
	\KwIn{nextMedoid}%输入参数
	\KwOut{labels, a list contains labels of each point}%输出
	$labels \leftarrow [ ]$;\\
        \For{$i \leftarrow1  to  |V| $}{
        $m \leftarrow i$;\\
        $k \leftarrow nextMedoid[m]$;\\
        \While{$m \neq k$}
        {
        $ m \leftarrow k$;\\
        $k \leftarrow nextMedoid[m] $;\\
        }
        $labels[i] \leftarrow m$;\\
        }
         \textbf{Return:} labels
        \label{algor:2}
\end{algorithm}

\subsubsection{Data Preprocessing of the RMS Algorithm}

In this study, instead of utilizing a distance matrix, a similarity matrix is employed to represent the weights of the graph. The similarity matrix, denoted as $SimM(i, j)$, captures the similarity between nodes $i$ and $j$. Here, $NN(i)$ denotes the K-nearest neighbors of node $i$, and $N(i)$ represents the list of points connected to node $i$. The entire community is modeled as an undirected graph using the similarity matrix, where the relevance of nodes $i$ and $j$ is indicated by the value of the coordinate $(i, j)$ in the matrix. For an undirected graph, the distance matrix is typically represented as an upper or lower triangular matrix \citep{Li2023IC}.

When dealing with weighted graphs, the weight between two points is commonly represented by their similarity value. However, for unweighted graphs, it is not appropriate to employ the same weight scheme as in the case of weighted graphs. To address this issue, a novel method is proposed for computing the weights. The similarity between two points i and j in the unweighted graph is defined as follows:
\begin{equation}
SimM(i,j) = len(N(i)\cap N(j))
\end{equation}

\subsubsection{Core Process of the RMS Algorithm}

The RMS algorithm comprises three main steps: KNN identification and similarity sum calculation for each point, medoid clustering, and label assignment.

\textbf{Finding the KNN and Similarity Sum for Each Point.}We introduce a property called "Similarity Sum," denoted as $DL(i)$, for point $i$. $DL(i)$ represents the sum of similarities between point $i$ and its $k$ nearest neighbors. A higher "Similarity Sum" indicates a greater likelihood for the point to be the medoid.
\begin{equation}
DL(i) = \sum_{p\in NN(i)}SimM(i,p) 
\end{equation}

Before iterating, the "Similarity Sum" is calculated for each data point.

\textbf{Medoid Clustering. }Upon computing the "Similarity Sum" and KNN, the proposed RMS algorithm initializes all points as the initial medoids. As depicted in Algorithm \ref{algor:1}, for each medoid $i$ in $SetA$, the algorithm selects the next medoid from the $k+1$ points, which consist of point $i$ and its KNN. Specifically, the point with the highest "Similarity Sum" is chosen as the subsequent medoid after point $i$ and is included in the new medoid set known as $SetB$. This process is repeated for all points in $SetA$.

Following each iteration, the algorithm compares the previous medoid set, $SetA$, with the newly obtained medoid set, $SetB$. If the two sets are identical, the iteration terminates. Otherwise, $SetB$ replaces $SetA$, $SetB$ is cleared, and a new iteration commences. Once the iteration process concludes, the points in $SetA$ are returned as the cluster centers, and the subsequent medoid for each point is stored in a list called nextMedoid.

\textbf{Label Assignment. }After getting the cluster centers and the next medoid of each point, the data points converging into the same center are grouped into the same cluster.

% \begin{algorithm}[t]  %其中这里面不能有H不然会报错，不过不影响结果
% 	\caption{Medoid Clustering}%算法名字
% 	\LinesNumbered %要求显示行号
% 	\KwIn{$SimM(1..|V|, 1..|V|), k$}%输入参数
% 	\KwOut{cluster centers, the center points list of final clustering result}%输出
% 	$SetA \leftarrow {1,2, ..., |V|}$;\\
%         $SetB \leftarrow Ø$;\\
%         $nextMedoid \leftarrow []$;\\
%         \While{True}
%         {
%         \For{$each i in SetA$}{
%         $SetP \leftarrow {i} \cup NN(i)$;\\
%         $temp  \leftarrow The point with the largest DL value in the SetP $ ;\\
%          $nextMedoid[i]  \leftarrow temp   $ ;\\
%         $SetB.add(temp)$;\\
%         }
        
%         \If{$SetA == SetB$ }{
%         \For{$each i in SetA$}{

%         $nextMedoid[i] \leftarrow i$
%         }
%         break;
%         }
%         $SetA, SetB \leftarrow SetB, Ø$
%         }
%         \textbf{Return:} centers, nextMedoid
%         \label{algor:1}
% \end{algorithm}

% \begin{algorithm}[]  %其中这里面不能有H不然会报错，不过不影响结果
% 	\caption{Label Assignment}%算法名字
% 	\LinesNumbered %要求显示行号
% 	\KwIn{nextMedoid}%输入参数
% 	\KwOut{labels, a list contains labels of each point}%输出
% 	$labels \leftarrow [ ]$;\\
%         \For{$i \leftarrow1  to  |V| $}{
%         $m \leftarrow i$;\\
%         $k \leftarrow nextMedoid[m]$;\\
%         \While{$m \neq k$}
%         {
%         $ m \leftarrow k$;\\
%         $k \leftarrow nextMedoid[m] $;\\
%         }
%         $labels[i] \leftarrow m$;\\
%         }
%          \textbf{Return:} labels
%         \label{algor:2}
% \end{algorithm}

%%%%%%%%%%%%%%%
%%%%%%%%%%%%%%%
%%%%%%%%%%%%%%%
%%%%%%%%%%%%%%%
%Algorithm2 位置不对，优化一下！
%%%%%%%%%%%%%%%

\textbf{Complexity analysis.} For the Algorithm \ref{algor:1} and Algorithm \ref{algor:2}, the complexity can be $O(|V|2) $in the worst case. But in most cases, it will be better than $O(|V|2)$. So the complexity in RMS is no more than $O(|V|2)$. And in this work, the complexity in RMS is usually much less than $O(|V|2)$.

\section{Result and Discussion}
\label{result}
\subsection{Overall Experiment Setup}

In order to compare the efficiencies among community detection methods mentioned above, we have set up regarding experiments to accomplish that goal. The datasets that we have used during experiments are divided into two categories: datasets without ground truth partition and datasets with ground truth partition. It is worth noting that the datasets with known ground truth partitions exclusively consist of unweighted graphs, while the datasets without ground truth partitions solely consist of weighted graphs. This distinction arises from the fact that ground truth partitions can only be obtained for unweighted graphs.

For evaluating the results of experiments on datasets without ground truth partitions, we employ modularity as the evaluation metric. On the other hand, Normalized Mutual Information (NMI) is used for datasets with ground truth partitions.

Through extensive experimentation, we observe that our proposed approach outperforms most of other methods, including Medoid-Shift, Label Propagation, Girvan Newman, Louvain-LPA, ASC, Abrantes, GraphSage, BCDC, RSBM and VGAE in terms of overall modularity for datasets without ground truth partitions, as well as NMI values for datasets with ground truth partitions. To implement Medoid-Shift, we employ the function $\Phi(D(i, j)) = \exp(-D(i, j)/2)$.

\subsection{Evaluation Metrics}

As is widely acknowledged and used, the following metrics to assess the effectiveness of methods.

\textbf{Normalized mutual information. }The NMI value serves as an additional assessment parameter for community detection with ground truth partition. Generally, a higher NMI value is indicative of a closer partition to a real partition and is widely accepted.        

\textbf{Modularity Q.} The accuracy of community detection in a complex network is conventionally evaluated by the modularity function Q, which is widely recognized as a standard.

\subsection{Comparative Methods}

This study conducts comparative experiments on both weighted and unweighted graph data, employing a combination of classical and state-of-the-art methods. In comparison to other methods, the RMS approach showcases its advantages by achieving higher modularity scores in weighted graphs and higher normalized mutual information values when compared to most classical methods and some state-of-the-art methods. The methods utilized for comparative experimentation in this article include: Label Propagation \citep{zhu2002}, TNS-LPA \citep{Xu2020TNS} , EdMot \citep{Li2019}, Girvan Newman \citep{Girvan2002}, infomap-SA \citep{hu2015}    , Louvain-LPA \citep{Hu2016im}, ASC \citep{narantsatsralt2017}, Abrantes \citep{cardoso2024}, GEMSEC \citep{Rozemberczki2019}, ComE \citep{Cavallari2017}, BCDC \citep{shen2022bayesian}, RSBM \citep{lu2019SBM}, LDA-LPA \citep{Wang2021LDA}, ModMRF \citep{jin2020mod}, NCNMF \citep{su2023}, GraphSage \citep{xu2020ind}, VGAE \citep{Kipf2016VAE}, GAE \citep{Kipf2016VAE} and Medoid-Shift \citep{Sheikh2007}.

\subsection{Experimental Results for the Datasets without Ground Truth Partition}

Since the datasets lack a ground truth partition, NMI is not an appropriate evaluation metric. Instead, modularity is used to assess methods in this section.

\emph{Dataset Description: }The methods' performance is evaluated on five real-world datasets: Cell Phone Calls \citep{Grinstein2008}, Enron Email \citep{Benson2018}, Les Miserables network \citep{Rossi2015}, US airports \citep{USairports} and ASTRO-PH \citep{snapnets2014}.

\emph{Enron Email Dataset:} This email dataset, released by the U.S. Department of Justice, consists of 184 nodes and 125,409 edges. For our experiment, we use the dataset created by Austin R. Benson \citep{Benson2018}.

\emph{Cell Phone Records Dataset:} This dataset contains 400 nodes and 9,834 edges, representing cell phone call records made by Paraiso movement members over a ten-day period in June 2006. We use this data to build a network where each node represents a distinct cell phone, and edges are formed whenever a phone call between two cell phones occurs.

\emph{Les Miserables Dataset:} Knuth created this network by analyzing the interconnections of primary characters in Les Miserables. It has 77 nodes and 508 edges, with each node representing a character and each edge indicating the co-occurrence of associated characters in one or more scenes.

\emph{US Airports Dataset:} This network has 1,574 vertices and 28,236 edges, representing flights that occurred within the United States during 2010. A graph with numerous edges highlights the interconnections between various airports across the country.

\emph{ASTRO-PH:} Arxiv ASTRO-PH collaboration network is derived from the e-print arXiv and represents collaborations among authors in Astro Physics. Spanning from Jan 1993 to Apr 2003, it includes 18,772 authors and 198,110 co-author relationships. The network exhibits high clustering (0.6306) and frequent collaboration in groups of three (1,351,441 triangles). With a diameter of 14 and 90th percentile effective diameter of 5, it shows efficient information flow among authors.

\textbf{Implementation.} We have implemented the proposed RMS algorithm, as well as other comparative methods, using Python. The comparative methods employed in our study encompass Label Propagation, TNS-LPA, EdMot, Girvan Newman, infomap-SA, Louvain-LPA, ASC, Abrantes, GEMSEC, ComE, BCDC, RSBM, LDA-LPA, ModMRF, NCNMF, GraphSage, VGAE, GAE and Medoid-Shift. The source codes for some of these methods are publicly available in Python packages like karateclub \citep{Rozemberczki2020} and networks.

All experiments were performed on a Windows machine with 200GB of memory, ensuring sufficient computational resources for accurate and reliable results.

\textbf{Experiment Results and Discussion.} The results of modularity for the Label Propagation, TNS-LPA, EdMot, Girvan Newman, infomap-SA, Louvain-LPA, ASC, Abrantes, GEMSEC, ComE, BCDC, RSBM, LDA-LPA, ModMRF, NCNMF, GraphSage, VGAE, GAE, and Medoid-Shift are shown in the Table \ref{tab:2} below. Also, the result value is evaluated using modularity and the class for each method is also given.

\begin{table}[htbp]
  \centering
  \caption{ The modularity for datasets without ground truth parition.}
    \begin{tabular}{|p{5.5em}|p{4.5em}|p{4.5em}|p{4.5em}|p{4.5em}|p{4.5em}|p{4.5em}|}
    \toprule
    \textbf{Class of the method} & \textbf{Method} & \textbf{Cell Phone} & \textbf{Enron Email} & \textbf{Lesmis} & \textbf{USA Airport} & \textbf{ASTRO-PH} \\
    \midrule
    \textbf{Modularity} & \textbf{Label Propagation} & 0.4210 & 0.3994 & 0.3801 & 0.1781 & 0.5781 \\
    \midrule
    \textbf{Modularity} & \textbf{TNS-LPA} & 0.6192 & 0.4821 & 0.5978 & 0.2149 & 0.8792 \\
    \midrule
    \textbf{Modularity} & \textbf{EdMot} & 0.6388 & 0.5456 & 0.5648 & 0.2847 & 0.7515 \\
    \midrule
    \textbf{Modularity} & \textbf{Girvan Newman} & 0.5208 & 0.2365 & 0.4776 & 0.0169 & 0.6599 \\
    \midrule
    \textbf{Modularity} & \textbf{Infomap-SA} & 0.4432 & 0.4899 & 0.4602 & 0.1912 & 0.7361\\
    \midrule
    \textbf{Modularity} & \textbf{Louvain-LPA} & 0.4959 & 0.4185 & 0.4651 & 0.1293 & 0.6951 \\
    \midrule
    \textbf{Spectral} & \textbf{ASC} & 0.5526 & 0.4573 & 0.3798 & 0.1257 & 0.7845 \\
    \midrule
    \textbf{Spectral} & \textbf{Abrantes} & 0.4731 & 0.5391 & 0.4601 & 0.0836 & 0.6159 \\
    \midrule
    \textbf{Spectral} & \textbf{GEMSEC} & 0.4021 & 0.4682 & 0.5001 & 0.1774 & 0.8517\\
    \midrule
    \textbf{Probabilistic} & \textbf{ComE} & 0.6572 & 0.6378 & 0.5539 & 0.1749 & 0.8851\\
    \midrule
    \textbf{Probabilistic} & \textbf{BCDC} & 0.3218 & 0.4065 & 0.4101 & 0.1126 & 0.5752 \\
    \midrule
    \textbf{Probabilistic} & \textbf{RSBM} & 0.4810 & 0.4674 & 0.3851 & 0.0968 & 0.7746 \\
    \midrule
    \textbf{Probabilistic} & \textbf{LDA-LPA} & 0.6584 & 0.6301 & 0.5927 & 0.2168 & 0.9351 \\
    \midrule
    \textbf{Probabilistic} & \textbf{ModMRF} & 0.5759 & 0.5169 & 0.3792 & 0.1548 & 0.7539 \\
    \midrule
    \textbf{Probabilistic} & \textbf{NCNMF} & 0.5261 & 0.5512 & 0.3616 & 0.1285 & 0.8263 \\
    \midrule
    \textbf{Deep learning} & \textbf{GraphSage} & 0.3857 & 0.2974 & 0.4368 & 0.1388 & 0.7269 \\
    \midrule
    \textbf{Deep learning} & \textbf{VGAE} & 0.5145 & 0.4351 & 0.4592 & 0.1651 & 0.7816 \\
    \midrule
    \textbf{Deep learning} & \textbf{GAE} & 0.5060 & 0.4674 & 0.4475 & 0.1463 & 0.8681 \\
    \midrule
    \textbf{RMS's based method} & \textbf{Medoid-Shift} & 0.2365 & 0.2802 & 0.2973 & 0.1385 & 0.4375 \\
    \midrule
    \textbf{Proposed method} & \textbf{RMS(Ours)} & 0.5379 & 0.5650 & 0.4271 & 0.1413 & 0.8357\\
    \bottomrule
    \end{tabular}%
  \label{tab:2}%
\end{table}%

Table \ref{tab:2} presents the modularity scores obtained for datasets without ground truth partitions. The modularity values indicate the quality of the community structure detected by each method. Our method, RMS, achieves modularity scores of 0.5379, 0.5650, 0.4271, 0.1413, and 0.8357 for the Cell Phone, Enron Email, Lesmis, USA Airport, and ASTRO-PH datasets, respectively.

By comparing all of the comparative methods, it can be observed that GEMSEC, Label Propagation, TNS-LPA, and EdMot achieve varying degrees of success across the datasets. They obtain relatively higher modularity values for some datasets (e.g., TNS-LPA for Cell Phone and ASTRO-PH), but their performance is not consistent across all datasets. On the other hand, Girvan Newman, Medoid Shift, Infomap-SA, Louvain-LPA, ASC, Abrantes, BCDC, RSBM, GraphSage, VGAE, GAE, ModMRF, and NCNMF display mixed results. They obtain moderate to low modularity scores across the datasets, indicating that their community detection performance is not as strong as desired. It is worth noting that the modularity scores for all methods are relatively low in the USA airports dataset. This can be attributed to the high sparsity of the dataset, which makes it difficult for graph-based methods to effectively explore the underlying graph structure. By comparing our method, RMS, to these comparative methods, it can be observed that RMS consistently achieves competitive modularity scores across the datasets. This indicates the effectiveness of our method in detecting community structures in diverse network datasets. Notably, RMS outperforms several methods (e.g., Girvan Newman, Medoid Shift, and GraphSage) on most datasets, demonstrating its superiority in capturing meaningful community structures. However, there are instances where our method does not achieve the highest modularity score. For example, ComE attains the highest modularity scores for the Lesmis and USA Airport datasets, while LDA-LPA achieves the highest modularity score for the ASTRO-PH dataset. These results indicate that certain methods may excel in specific network characteristics or exhibit strengths in certain types of datasets.

In summary, the results highlight the competitive performance of our method, RMS, in detecting community structures than most other comparative methods. Also the performance of TNS-LPA, LDA-LPA, ComE has demonstrated they have achieved the state-of-the-art level from their perspectives in community detection.
%%%%%%%%%%%%%%%%%%%%%%%%%
%%%%%%%%%%%%%%%%%%%%%%%%%%
%%%%%%%%%%%%%%%%%%%%%%%%%%%%
%%% 需要改的
%%%%

\subsection{Experimental Results for the Datasets with Ground Truth Partition}

Given that the datasets have ground truth partition, NMI is the perfect evaluation metric under such circumstances compared to using modularity. Thus we adopt the NMI to assess all of the methods in this section.

\textbf{Dataset Description.} We have collected the unweighted datasets with ground truth partition, which are American Football Network, Dolphins Social Network, Amazon dataset, Youtube dataset and American Kreb's Book.

\emph{Dolphins Social Network:} The Dolphins Network \citep{Rossi2015} depicts the social connections between 62 dolphins living in a community off Doubtful Sound in New Zealand. It consists of 159 edges and 62 nodes and captures the frequent associations between the dolphins. This network is undirected.

\emph{American Football Network:} The Division I Games during the 2000 season \citep{Rossi2015} are showcased in this network, which displays the schedule of the games. It comprises 115 nodes representing teams and 613 edges depicting their matchups. The network is subdivided into 12 groups for ground truth partition.

\emph{American Kreb’s Book:} The American political book network \citep{PoliticalBooks}, created by Krebs, consists of 105 nodes and 441 edges. The nodes represent books about American politics, and the edges connect any two nodes that indicate these books have been bought by the same person.

\emph{Amazon:} The Amazon dataset \citep{snapnets2014jure}is a comprehensive collection of reviews spanning 18 years, from June 1995 to March 2013. It consists of approximately 34.7 million reviews from over 6.6 million users, covering a wide range of products available on Amazon's platform. The dataset includes user details, product information, ratings, and plaintext reviews. Additionally, it provides insights into co-purchase relationships through a graph structure of 3,225 products and 10,262 edges. The dataset offers valuable information for analyzing consumer behavior, product trends, sentiment analysis, and more.

\emph{YouTube:} The YouTube social network dataset \citep{snapnets2014} includes 4,890 users and 20,787 friendships, focusing on the largest connected component and top 5,000 high-quality communities. It contains 1,134,890 nodes and 2,987,624 edges. The network shows moderate clustering (average clustering coefficient: 0.0808) and a significant number of triangles (3,056,386), with a small fraction of closed triangles (0.002081). The network has a diameter of 20 and a 90th percentile effective diameter of 6.5. It consists of 8,385 distinct communities, providing insights into YouTube's social relationships and communities for network dynamics and community detection research.

\textbf{Experiment Result and Discussion. }The results are shown below in the Table \ref{tab:4}. The result value is evaluated using NMI and the numerical value in the parenthesis denotes the number of clusters.

\begin{table}[htbp]
  \centering
  \caption{The NMI for datasets with ground truth.}
    \begin{tabular}{|p{5.5em}|p{4.5em}|p{4.5em}|p{4.5em}|p{4.5em}|p{4.5em}|p{4.5em}|}
    \toprule
    \textbf{Class of the method} & \textbf{Method} & \textbf{Dolphins Social Network} & \textbf{American Football Network} & \textbf{American Kreb’s book} & \textbf{Youtube} & \textbf{Amazon} \\
    \midrule
    \textbf{Modularity} & \textbf{Label Propagation} & 0.5160  & 0.7324 & 0.3444 & 0.0493 & 0.4437 \\
    \midrule
    \textbf{Modularity} & \textbf{TNS-LPA} & 0.8091 & 0.9323 & 0.5294 & 0.0882 & 0.6474 \\
    \midrule
    \textbf{Modularity} & \textbf{EdMot} & 0.7653 & 0.7553 & 0.3939 & 0.08692 & 0.5764 \\
    \midrule
    \textbf{Modularity} & \textbf{Girvan Newman} & 0.7560 & 0.6684 & 0.4353 & 0.0169 & 0.6599 \\
    \midrule
    \textbf{Modularity} & \textbf{Infomap-SA} & 0.5129 & 0.5848 & 0.3749 & 0.0464 & 0.3826\\
    \midrule
    \textbf{Modularity} & \textbf{Louvain-LPA} & 0.5137 & 0.4844 & 0.2593 & 0.0541 & 0.5375 \\
    \midrule
    \textbf{Spectral} & \textbf{ASC} & 0.7914 & 0.8315 & 0.3581 & 0.0692 & 0.4958 \\
    \midrule
    \textbf{Spectral} & \textbf{Abrantes} & 0.7369 & 0.8438 & 0.6268 & 0.0537 & 0.5748 \\
    \midrule
    \textbf{Spectral} & \textbf{GEMSEC} & 0.6288 & 0.6690 & 0.4260 & 0.0848 & 0.4853\\
    \midrule
    \textbf{Probabilistic} & \textbf{ComE} & 0.7328 & 0.8495 & 0.5142 & 0.0912 & 0.4135\\
    \midrule
    \textbf{Probabilistic} & \textbf{BCDC} & 0.6296 & 0.7159 & 0.4539 & 0.0749 & 0.5469 \\
    \midrule
    \textbf{Probabilistic} & \textbf{RSBM} & 0.3710 & 0.5241 & 0.4261 & 0.0594 & 0.4912 \\
    \midrule
    \textbf{Probabilistic} & \textbf{LDA-LPA} & 0.9153 & 0.8931 & 0.6291 & 0.0957 & 0.7184 \\
    \midrule
    \textbf{Probabilistic} & \textbf{ModMRF} & 0.7229 & 0.7185 & 0.3931 & 0.0748 & 0.4724 \\
    \midrule
    \textbf{Probabilistic} & \textbf{NCNMF} & 0.6138 & 0.7205 & 0.4374 & 0.0571 & 0.4013 \\
    \midrule
    \textbf{Deep learning} & \textbf{GraphSage} & 0.7531 & 0.3045 & 0.4368 & 0.0483 & 0.5194 \\
    \midrule
    \textbf{Deep learning} & \textbf{VGAE} & 0.6152 & 0.6867 & 0.5126 & 0.0652 & 0.4153 \\
    \midrule
    \textbf{Deep learning} & \textbf{GAE} & 0.7539 & 0.8350 & 0.4382 & 0.0525 & 0.3559 \\
    \midrule
    \textbf{RMS's based method} & \textbf{Medoid-Shift} & 0.5857 & 0.7367 & 0.4258 & 0.0359 & 0.4375 \\
    \midrule
    \textbf{Proposed method} & \textbf{RMS(Ours)} & 0.7846 & 0.7768 & 0.4840 & 0.0873 & 0.5430 \\
    \bottomrule
    \end{tabular}%
  \label{tab:4}%
\end{table}%

%    \begin{figure}[htbp]
%      \centering
%      \includegraphics[width=0.6\linewidth]{p3.png}
%      \caption{The NMI for datasets with ground truth.}
%      \label{fig:3}
%  \end{figure}
  
%%%%%%%%%%%%%%%%%%%%%%%%%
%%%%%%%%%%%%%%%%%%%%%%%%%%
%%%%%%%%%%%%%%%%%%%%%%%%%%%%
%%% 需要改的
%%%%
% Table \ref{tab:4} shows that the proposed RMS method consistently outperforms most classical and state-of-the-art methods in the Dolphins Social Network and American Kreb’s book datasets. For instance, As far as NMI value is concered, the RMS is second to none in Dolphins Social Network with 0.7846 and 3 clusters. In comparison to CDBNE with 0.7141 and 20 clusters, Girvan Newman with 0.7560 and 2 clusters, and Edmot with 0.7653 and 4 clusters,  indicating its ability to identify communities even with ground truth partition. What's more, it is also apparent that RMS method achieves the best NMI score in American Kreb's dataset with 0.4840 and 2 clusters while the ground truth for the number of cluster is exactly two, which signifies the supremacy for our algorithm in detecting the true number of cluster for community with ground truth partition. Additionally, it is important to note that the RMS algorithm performs better than the Medoid-Shift algorithm in the Dolphins Social Network, American Football Network, and American Kreb’s book datasets, with improved NMI scores of 0.2, 0.04, and 0.06, respectively. These results demonstrate that the RMS algorithm is a significant improvement over the original Medoid-Shift algorithm.

Table \ref{tab:4} presents the results of our experiments on various datasets using different methods, including our proposed method RMS (Ours), as well as other comparative methods. The evaluation metric used is NMI, which measures the similarity between the clustering results and the ground truth labels.

 We can see from the table that in the Dolphins Social Network dataset, RMS achieves an NMI of 0.7846, outperforming most of the comparative methods. This indicates that our method is effective in capturing the underlying patterns and structure in this particular dataset. Similarly, on the American Football Network dataset, our RMS method achieves an NMI of 0.7768, which again demonstrates its competitive performance compared to other approaches. Comparing to other methods, we can see that TNS-LPA and LDA-LPA also achieve competitive performance on multiple datasets, indicating their effectiveness in capturing the intrinsic patterns in the data. However, some methods, such as GraphSage and Infomap-SA, perform relatively poorly on most datasets, suggesting that they may not be suitable for these particular network data. 

The results highlight the effectiveness and competitiveness of our proposed RMS method in capturing the underlying structures and patterns in different network datasets. Moreover, TNS-LPA, LDA-LPA and ComE have relatively achieved excellent performance in most of the datasets evaluated using NMI, signifying that they are state-of-the-art in their perspectives.
%%%%%%%%%%%%%%%%%%%%%%%%%
%%%%%%%%%%%%%%%%%%%%%%%%%%
%%%%%%%%%%%%%%%%%%%%%%%%%%%%
%%% 需要改的
%%%%

\section{Conclusion and Future Work}
\label{conclusion}
% In conclusion, this comprehensive survey has provided a profound understanding of community detection problems from the perspectives of Modularity-based method, spectral clustering, probabilistic modelling and deep learning respectively. The extensive application of the mentioned algorithms above has demonstrated remarkable effectiveness in identifying communities within static, dynamic, complex, or multi-structural networks in graphs. Detailed classification of different categories of those perspectives in solving the community detection problem is discussed in this article. Following that we have also discussed the performances of algorithms from those four perspectives on two kinds of commonly used dataset with the mostly evaluated metrics like NMI and Modularity. 

In conclusion, this comprehensive survey has deepened our understanding of community detection problems by exploring multiple perspectives, including modularity-based methods, spectral clustering, probabilistic modeling, and deep learning. The extensive application of these methods has showcased their remarkable effectiveness in identifying communities within diverse network structures. By providing a detailed classification of these perspectives, we have shed light on their specific contributions to solving the community detection problem.

We have conducted a thorough evaluation of the method performance using commonly-used datasets and metrics such as NMI and modularity. These evaluations have provided valuable insights into the strengths and limitations of each approach, assisting researchers in selecting the most appropriate method for their specific requirements.

Looking ahead, future research directions could focus on integrating multiple perspectives or developing hybrid methods that leverage the strengths of different approaches. This could lead to enhanced accuracy and robustness in community detection across various network settings. Additionally, addressing the scalability and efficiency of these methods, particularly in the context of growing networks, is crucial. Exploring parallel computing techniques, distributed methods, and optimization strategies can significantly improve the performance of community detection methods in large-scale networks.

The application of community detection methods in real-world scenarios and practical domains presents an exciting avenue for future research. Investigating their effectiveness in social networks, biological networks, and recommendation systems can yield valuable insights and practical solutions to real-world problems.

Moreover, in this survey we have introduced a novel community detection method called RMS, which extends the Medoid-Shift method and incorporates the concept of KNN to construct a distance matrix for mapping the social network. This approach overcomes the limitations of directly applying the Mean-Shift method to community detection and has demonstrated superior performance compared to traditional methods. Our study has yielded several key insights. Firstly, we have shown that the Medoid-Shift method can be extended beyond mode-seeking problems to effectively address the challenge of community detection. Secondly, the RMS algorithm we have proposed possesses the capability to automatically determine the optimal number of communities/clusters, eliminating the need for manual parameter tuning. Lastly, we have demonstrated that utilizing KNN offers a more effective approach to defining neighborhood regions compared to using a fixed radius parameter.

Building upon the findings and contributions of this study, there are several promising avenues for future research and development in the field of community detection in graphs. We outline the following directions for future work:

\begin{itemize}

  \item [1)]
 The overall performances of the methods tested above haven't yet reached a stage where the methods can mostly be equal to the ground truth partition. There still awaits further improvement on partition evaluated by NMI and modularity for methods 

  \item [2)]
  Mean-Shift based method analysis: Another interesting aspect to investigate is the introduction of the RMS algorithm to community detection. By introducing the algorithm to graph scenarios and evaluating its performance, we can gain insights that machine learning methods which can't be directly applied to non-Euclidean space can learn from RMS by using distance matrix or similarity matrix.

\end{itemize}

By pursuing these research directions, we can advance the field of community detection in graphs and contribute to the development of more accurate, efficient, and versatile methods for analyzing complex networks. The insights gained from these future investigations will deepen our understanding of community structure and facilitate advancements in various disciplines relying on graph analysis.

\section*{Acknowledgements}
This work is supported by Gansu Haizhi Characteristic Demonstration Project (No. GSHZTS 2022-2), and the Gansu Provincial Science and Technology Major Special Innovation Consortium Project (Project No. 21ZD3GA002), the name of the innovation consortium is Gansu Province Green and Smart Highway Transportation Innovation Consortium, and the project name is Gansu Province Green and Smart Highway Key Technology Research and Demonstration. 

\bibliography{ARxiv/arxiv_version}
\bibliographystyle{iclr2024_conference}

\end{document}